\RequirePackage{lineno}
\documentclass[aps,prd,twocolumn,showpacs,superscriptaddress,groupedaddress,amsmath,amssymb]{revtex4}

\usepackage{graphicx}
\usepackage{graphics}
\usepackage{dcolumn}
\usepackage{bm}
\usepackage{amssymb} 
\usepackage{multirow} 

\newcommand {\Bd} {\ensuremath{B^0}}
\newcommand {\Bs} {\ensuremath{B^0_s}}
\newcommand {\Bq} {\ensuremath{B^0_q}}

\newcommand {\barBq} {\ensuremath{\bar{B}^0_q}}
\newcommand {\asld} {\ensuremath{a^d_{\mathrm{sl}}}}
\newcommand {\asls} {\ensuremath{a^s_{\mathrm{sl}}}}
\newcommand {\aslq} {\ensuremath{a^q_{\mathrm{sl}}}}
\newcommand {\aslb} {\ensuremath{A^b_{\mathrm{sl}}}}
\newcommand {\ktomu} {\ensuremath{K \to \mu}}
\newcommand {\pitomu} {\ensuremath{\pi \to \mu}}
\newcommand {\ptomu} {\ensuremath{p \to \mu}}
\newcommand {\ks} {\ensuremath{K^0_S}}

\begin{document}

\hspace{5.2in} \mbox{Fermilab-Pub-11/307-E}

\title{
Measurement of the anomalous like-sign dimuon charge asymmetry
with 9 fb$\bm{^{-1}}$ of $\bm{p \bar p}$~collisions
}

\affiliation{Universidad de Buenos Aires, Buenos Aires, Argentina}
\affiliation{LAFEX, Centro Brasileiro de Pesquisas F{\'\i}sicas, Rio de Janeiro, Brazil}
\affiliation{Universidade do Estado do Rio de Janeiro, Rio de Janeiro, Brazil}
\affiliation{Universidade Federal do ABC, Santo Andr\'e, Brazil}
\affiliation{Instituto de F\'{\i}sica Te\'orica, Universidade Estadual Paulista, S\~ao Paulo, Brazil}
\affiliation{Simon Fraser University, Vancouver, British Columbia, and York University, Toronto, Ontario, Canada}
\affiliation{University of Science and Technology of China, Hefei, People's Republic of China}
\affiliation{Universidad de los Andes, Bogot\'{a}, Colombia}
\affiliation{Charles University, Faculty of Mathematics and Physics, Center for Particle Physics, Prague, Czech Republic}
\affiliation{Czech Technical University in Prague, Prague, Czech Republic}
\affiliation{Center for Particle Physics, Institute of Physics, Academy of Sciences of the Czech Republic, Prague, Czech Republic}
\affiliation{Universidad San Francisco de Quito, Quito, Ecuador}
\affiliation{LPC, Universit\'e Blaise Pascal, CNRS/IN2P3, Clermont, France}
\affiliation{LPSC, Universit\'e Joseph Fourier Grenoble 1, CNRS/IN2P3, Institut National Polytechnique de Grenoble, Grenoble, France}
\affiliation{CPPM, Aix-Marseille Universit\'e, CNRS/IN2P3, Marseille, France}
\affiliation{LAL, Universit\'e Paris-Sud, CNRS/IN2P3, Orsay, France}
\affiliation{LPNHE, Universit\'es Paris VI and VII, CNRS/IN2P3, Paris, France}
\affiliation{CEA, Irfu, SPP, Saclay, France}
\affiliation{IPHC, Universit\'e de Strasbourg, CNRS/IN2P3, Strasbourg, France}
\affiliation{IPNL, Universit\'e Lyon 1, CNRS/IN2P3, Villeurbanne, France and Universit\'e de Lyon, Lyon, France}
\affiliation{III. Physikalisches Institut A, RWTH Aachen University, Aachen, Germany}
\affiliation{Physikalisches Institut, Universit{\"a}t Freiburg, Freiburg, Germany}
\affiliation{II. Physikalisches Institut, Georg-August-Universit{\"a}t G\"ottingen, G\"ottingen, Germany}
\affiliation{Institut f{\"u}r Physik, Universit{\"a}t Mainz, Mainz, Germany}
\affiliation{Ludwig-Maximilians-Universit{\"a}t M{\"u}nchen, M{\"u}nchen, Germany}
\affiliation{Fachbereich Physik, Bergische Universit{\"a}t Wuppertal, Wuppertal, Germany}
\affiliation{Panjab University, Chandigarh, India}
\affiliation{Delhi University, Delhi, India}
\affiliation{Tata Institute of Fundamental Research, Mumbai, India}
\affiliation{University College Dublin, Dublin, Ireland}
\affiliation{Korea Detector Laboratory, Korea University, Seoul, Korea}
\affiliation{CINVESTAV, Mexico City, Mexico}
\affiliation{Nikhef, Science Park, Amsterdam, the Netherlands}
\affiliation{Radboud University Nijmegen, Nijmegen, the Netherlands and Nikhef, Science Park, Amsterdam, the Netherlands}
\affiliation{Joint Institute for Nuclear Research, Dubna, Russia}
\affiliation{Institute for Theoretical and Experimental Physics, Moscow, Russia}
\affiliation{Moscow State University, Moscow, Russia}
\affiliation{Institute for High Energy Physics, Protvino, Russia}
\affiliation{Petersburg Nuclear Physics Institute, St. Petersburg, Russia}
\affiliation{Instituci\'{o} Catalana de Recerca i Estudis Avan\c{c}ats (ICREA) and Institut de F\'{i}sica d'Altes Energies (IFAE), Barcelona, Spain}
\affiliation{Stockholm University, Stockholm and Uppsala University, Uppsala, Sweden}
\affiliation{Lancaster University, Lancaster LA1 4YB, United Kingdom}
\affiliation{Imperial College London, London SW7 2AZ, United Kingdom}
\affiliation{The University of Manchester, Manchester M13 9PL, United Kingdom}
\affiliation{University of Arizona, Tucson, Arizona 85721, USA}
\affiliation{University of California Riverside, Riverside, California 92521, USA}
\affiliation{Florida State University, Tallahassee, Florida 32306, USA}
\affiliation{Fermi National Accelerator Laboratory, Batavia, Illinois 60510, USA}
\affiliation{University of Illinois at Chicago, Chicago, Illinois 60607, USA}
\affiliation{Northern Illinois University, DeKalb, Illinois 60115, USA}
\affiliation{Northwestern University, Evanston, Illinois 60208, USA}
\affiliation{Indiana University, Bloomington, Indiana 47405, USA}
\affiliation{Purdue University Calumet, Hammond, Indiana 46323, USA}
\affiliation{University of Notre Dame, Notre Dame, Indiana 46556, USA}
\affiliation{Iowa State University, Ames, Iowa 50011, USA}
\affiliation{University of Kansas, Lawrence, Kansas 66045, USA}
\affiliation{Kansas State University, Manhattan, Kansas 66506, USA}
\affiliation{Louisiana Tech University, Ruston, Louisiana 71272, USA}
\affiliation{Boston University, Boston, Massachusetts 02215, USA}
\affiliation{Northeastern University, Boston, Massachusetts 02115, USA}
\affiliation{University of Michigan, Ann Arbor, Michigan 48109, USA}
\affiliation{Michigan State University, East Lansing, Michigan 48824, USA}
\affiliation{University of Mississippi, University, Mississippi 38677, USA}
\affiliation{University of Nebraska, Lincoln, Nebraska 68588, USA}
\affiliation{Rutgers University, Piscataway, New Jersey 08855, USA}
\affiliation{Princeton University, Princeton, New Jersey 08544, USA}
\affiliation{State University of New York, Buffalo, New York 14260, USA}
\affiliation{Columbia University, New York, New York 10027, USA}
\affiliation{University of Rochester, Rochester, New York 14627, USA}
\affiliation{State University of New York, Stony Brook, New York 11794, USA}
\affiliation{Brookhaven National Laboratory, Upton, New York 11973, USA}
\affiliation{Langston University, Langston, Oklahoma 73050, USA}
\affiliation{University of Oklahoma, Norman, Oklahoma 73019, USA}
\affiliation{Oklahoma State University, Stillwater, Oklahoma 74078, USA}
\affiliation{Brown University, Providence, Rhode Island 02912, USA}
\affiliation{University of Texas, Arlington, Texas 76019, USA}
\affiliation{Southern Methodist University, Dallas, Texas 75275, USA}
\affiliation{Rice University, Houston, Texas 77005, USA}
\affiliation{University of Virginia, Charlottesville, Virginia 22901, USA}
\affiliation{University of Washington, Seattle, Washington 98195, USA}
\author{V.M.~Abazov} \affiliation{Joint Institute for Nuclear Research, Dubna, Russia}
\author{B.~Abbott} \affiliation{University of Oklahoma, Norman, Oklahoma 73019, USA}
\author{B.S.~Acharya} \affiliation{Tata Institute of Fundamental Research, Mumbai, India}
\author{M.~Adams} \affiliation{University of Illinois at Chicago, Chicago, Illinois 60607, USA}
\author{T.~Adams} \affiliation{Florida State University, Tallahassee, Florida 32306, USA}
\author{G.D.~Alexeev} \affiliation{Joint Institute for Nuclear Research, Dubna, Russia}
\author{G.~Alkhazov} \affiliation{Petersburg Nuclear Physics Institute, St. Petersburg, Russia}
\author{A.~Alton$^{a}$} \affiliation{University of Michigan, Ann Arbor, Michigan 48109, USA}
\author{G.~Alverson} \affiliation{Northeastern University, Boston, Massachusetts 02115, USA}
\author{G.A.~Alves} \affiliation{LAFEX, Centro Brasileiro de Pesquisas F{\'\i}sicas, Rio de Janeiro, Brazil}
% ...
%
%
\collaboration{The D0 Collaboration\footnote{with visitors from
$^{a}$Augustana College, Sioux Falls, SD, USA,
$^{b}$The University of Liverpool, Liverpool, UK,
$^{c}$SLAC, Menlo Park, CA, USA,
$^{d}$University College London, London, UK,
$^{e}$Centro de Investigacion en Computacion - IPN, Mexico City, Mexico,
$^{f}$ECFM, Universidad Autonoma de Sinaloa, Culiac\'an, Mexico,
and
$^{g}$Universit{\"a}t Bern, Bern, Switzerland.
$^{\ddag}$Deceased.
}} \noaffiliation
\vskip 0.25cm

\date{June 30, 2011}

\begin{abstract}

\noindent
We present an updated measurement of the anomalous like-sign dimuon charge asymmetry $\aslb$
for semi-leptonic $b$-hadron decays in 9.0~fb$^{-1}$ of $p\overline{p}$ collisions
recorded with the D0 detector at a center-of-mass energy of $\sqrt{s} = 1.96$ TeV
at the Fermilab Tevatron collider. We obtain
$\aslb = (-0.787 \pm 0.172~({\rm stat}) \pm 0.093~({\rm syst})) \%$.
This result differs by $3.9$ standard deviations from the prediction of the standard model 
and provides evidence for anomalously large $CP$ violation in semi-leptonic
neutral $B$ decay. The dependence of the asymmetry on the muon
impact parameter is consistent with the hypothesis that
it originates from semi-leptonic $b$-hadron decays.


\end{abstract}


\maketitle



\section{Introduction}
\label{introduction}

We measure the like-sign dimuon charge asymmetry of semi-leptonic
decays of $b$ hadrons,
\begin{equation}
\aslb \equiv \frac{N^{++}_{b} - N^{--}_{b}}{N^{++}_{b} + N^{--}_{b}},
\end{equation}
in 9.0~fb$^{-1}$ of $p\bar{p}$ collisions recorded with the D0 detector at a center-of-mass
energy $\sqrt{s}=1.96$~TeV at the Fermilab Tevatron collider. Here
$N^{++}_{b}$ and $N^{--}_{b}$
are the number of events containing two positively charged
or two negatively charged muons, respectively, both of which are produced in
prompt semi-leptonic $b$-hadron decays.
At the Fermilab Tevatron $p \bar p$ collider, $b$ quarks are produced mainly in $b \bar{b}$ pairs.
Hence, to observe an event with two like-sign muons
from semi-leptonic $b$-hadron decay, one
of the hadrons must be a $\Bd$ or $\Bs$ meson that oscillates and decays to
a muon of charge opposite of that expected from the original $b$ quark \cite{charge}.
The oscillation $\Bq \leftrightarrow \barBq$ ($q = d$ or $s$) is described by
higher order loop diagrams
that are sensitive to hypothetical particles that may not be directly accessible at the Tevatron.

The asymmetry $\aslb$ has contributions from the semi-leptonic charge asymmetries
$\asld$ and $\asls$ of $\Bd$ and $\Bs$ mesons \cite{Grossman}, respectively:
\begin{eqnarray}
\aslb & = & C_d \asld + C_s \asls, \label{Ab_7} \\
\mathrm{with}\thickspace\aslq & = & \frac{\Delta \Gamma_q}{\Delta M_q} \tan \phi_q, \label{i_phiq}
\end{eqnarray}
where $\phi_q$ is a {\sl CP}-violating phase, and $\Delta M_q$ and
$\Delta \Gamma_q$ are the mass and width differences between the
eigenstates of the propagation matrices of the neutral $\Bq$ mesons.
The coefficients $C_d$ and $C_s$ depend
on the mean mixing probability, $\chi_0$, and the production rates of $\Bd$ and $\Bs$ mesons.
We use the values of these quantities measured at LEP as averaged by 
the Heavy Flavor Averaging Group (HFAG) \cite{hfag} and obtain
\begin{eqnarray}
C_d & = & 0.594 \pm 0.022, \nonumber \\
C_s & = & 0.406 \pm 0.022.
\label{Ab_8}
\end{eqnarray}
The value of $\chi_0$
measured by the CDF Collaboration recently \cite{cdf-chi0} is consistent with the LEP value,
which supports this choice of parameters.
Using the standard model (SM) prediction for $\asld$ and $\asls$ \cite{Nierste}, we find 
\begin{equation}
\aslb({\rm SM}) = (-0.028^{+0.005}_{-0.006}) \%,
\label{in_aslbsm}
\end{equation}
which is negligible compared to present experimental sensitivity.
Additional contributions to $CP$ violation via loop diagrams
appear in some extensions of the SM and can result in an asymmetry
$\aslb$ within experimental reach
\cite{Randall,Hewett,Hou,Soni,buras}.

This Article is an update to Ref.~\cite{PRD}
that reported evidence for an anomalous like-sign dimuon charge asymmetry with
6.1 fb$^{-1}$ of data, at the 3.2 standard deviation level.
All notations used here are given in Ref.~\cite{PRD}.
This new measurement is based on a larger dataset and further
improvements in the measurement technique. In addition, the asymmetry's dependence
on the muon impact parameter (IP)~\cite{impact} is studied. The D0 detector is described
in Ref.~\cite{d0-det}.
We include a brief overview of the
analysis in Sec.~\ref{overview}.
Improvements made to muon selections are presented in Sec.~\ref{selection};
the measurement of all quantities required to determine the asymmetry $\aslb$
is described in Secs.~\ref{sec_fk}--\ref{sec_Ab}, and the result is given
in Sec.~\ref{sec_ah}. Sections~\ref{sec_consistency}--\ref{sec_sim} present consistency checks of
the measurement; Sec.~\ref{sec_asip} describes the study of the asymmetry's IP dependence.
Conclusions are given in Sec. \ref{conclusions}.

\section{Method}
\label{overview}
The elements of our analysis are described in detail in Ref.~\cite{PRD}. Here,
we summarize briefly
the method, emphasizing the improvements to our previous procedure.
We use two sets of data:  \textit{(i) inclusive muon} data
collected with inclusive muon triggers that provide $n^+$ positively charged muons and
$n^-$ negatively charged muons, and \textit{(ii) like-sign dimuon} data,
collected with dimuon triggers that provide
$N^{++}$ events with two
positively charged muons and $N^{--}$ events with two negatively charged muons.
If an event contains more than one muon, each muon is included in the inclusive muon sample.
Such events constitute about 0.5\% of the total inclusive muon sample.
If an event contains more than two muons, the two muons with the highest transverse momentum
($p_T$) are selected for inclusion in the dimuon sample.
Such events comprise about 0.7\% of the total like-sign dimuon sample.

From these data we obtain
the inclusive muon charge asymmetry $a$ and the like-sign dimuon charge
asymmetry $A$, defined as
\begin{eqnarray}
a & = & \frac{n^+ - n^-}{n^+ + n^-}, \nonumber \\
A & = & \frac{N^{++} - N^{--}}{N^{++} + N^{--}}.
\end{eqnarray}
In addition to a possible signal asymmetry $\aslb$, these asymmetries have
contributions from muons produced in kaon and pion decay, or from hadrons that punch through
the calorimeter and iron toroids to penetrate the outer muon detector.
The charge asymmetry related
to muon detection and identification also contributes to $a$ and $A$. These contributions
are measured with data, with only minimal input from simulation. The largest contribution by far is from
kaon decays. Positively charged kaons have smaller cross sections in the detector material
than negatively charged kaons \cite{pdg}, giving them more time to decay. This difference
produces a positive charge asymmetry.

We consider muon candidates with $p_T$ in the range 1.5 to $25$~GeV.
This range is divided into six bins as shown in Table \ref{tab7}. The inclusive muon charge
asymmetry $a$
can be expressed \cite{PRD} as
\begin{equation}
a = \sum_{i=1}^6 f_{\mu}^i \{f^i_S (a_S + \delta_i) + f^i_K a^i_K + f^i_\pi a^i_\pi + f^i_p a^i_p\},
\label{inclusive_mu_a}
\end{equation}
where the fraction of reconstructed muons, $f_{\mu}^i$, in a given $p_T$ interval $i$
in the inclusive muon sample is given in Table \ref{tab7}.
The fractions of these muons produced by kaons, pions, and protons in a given
$p_T$ interval $i$ are $f^i_K$, $f^i_\pi$, and $f^i_p$,
and their charge asymmetries are $a^i_K$, $a^i_\pi$, and $a^i_p$, respectively.
We refer to
these muons as ``long" or ``$L$" muons since they are produced by particles traveling long
distances before decaying within the detector material. The track of a $L$ muon in the central
tracker is dominantly produced by the parent hadron. The charge asymmetry of $L$ muons results from
the difference in the interactions of positively and negatively charged particles with the detector material,
and is not related to {\sl CP} violation. The background fraction is defined as
$f_{\rm{bkg}}^i = f^i_K + f^i_\pi + f^i_p$. The quantity $f_S^i = 1 - f_{\rm{bkg}}^i$ is
the fraction of muons from weak decays of $b$ and $c$ quarks and $\tau$ leptons, and from
decays of short-lived mesons ($\phi, \omega, \eta, \rho^0$). We refer to
these muons as ``short" or ``$S$" muons, since they arise from the decay of particles at small
distances from the $p \bar p$ interaction point. These particles are not affected by interactions in
the detector material, and
once muon detection and identification
imbalances are removed,
the muon charge asymmetry $a_S$ must therefore be produced only through
{\sl CP} violation in the underlying physical processes.
The quantity $\delta_i$ in Eq.~(\ref{inclusive_mu_a}) is the charge asymmetry related
to muon detection and identification.
The background charge asymmetries $a^i_K$, $a^i_\pi$, and $a^i_p$ are measured in the
inclusive muon data, and include any detector asymmetry. The
$\delta_i$ therefore accounts only for $S$ muons and is multiplied by the factor $f^i_S$.

The like-sign dimuon charge asymmetry $A$
can be expressed \cite{PRD} as
\begin{eqnarray}
A & = & F_{SS} A_S + F_{SL} a_S + \sum_{i=1}^6 F^i_\mu \{(2 - F^i_{\rm bkg}) \delta_i \nonumber \\
  &   & + F^i_K a^i_K + F^i_\pi a^i_\pi + F^i_p a^i_p \}.
\label{dimuon_A}
\end{eqnarray}
The quantity $A_S$ is the charge asymmetry of the events with two like-sign $S$ muons.
The quantity $F_{SS}$ is the fraction of like-sign dimuon events
with two $S$ muons, $F_{SL}$ is the fraction of like-sign dimuon events
with one $S$ and one $L$ muon. We also define the quantity $F_{LL}$ as
the fraction of like-sign dimuon events
with two $L$ muons.
The quantity $F^i_\mu$ is the fraction of muons in the $p_T$ interval $i$
in the like-sign dimuon data. The quantities
$F^i_x$ ($x= K, \pi, p$) are defined as $F^i_x \equiv 2 N_x^i/N_{\mu}^i$, where
$N_x^i$ is the number of muons produced by kaons, pions, and protons, respectively,
in a $p_T$ interval $i$, with $N_{\mu}^i$ being the number of muons in this interval, with
the factor of two taking into account the normalization of these quantities per like-sign dimuon event.
The quantity $F_{\rm bkg}^i$ is a sum over muons produced by hadrons:
\begin{equation}
F_{\rm bkg}^i  \equiv F_K^i + F_\pi^i + F_p^i.
\end{equation}
We also define $F_{\rm bkg}$ as
\begin{eqnarray}
F_{\rm bkg} & \equiv & \sum_{i=1}^6 (F^i_\mu F^i_{\rm bkg}) \\
            & =      & F_{SL} + 2 F_{LL} \nonumber \\
            & =      & 1 + F_{LL} - F_{SS}.
\end{eqnarray}
The estimated contribution from the neglected quadratic terms in Eq.~(\ref{dimuon_A})
is approximately $2 \times 10^{-5}$, which corresponds to about 5\% of the statistical
uncertainty on $A$.

The asymmetries $a_S$ and $A_S$ in Eqs.~(\ref{inclusive_mu_a}) and (\ref{dimuon_A})
are the only asymmetries due to {\sl CP} violation
in the processes producing $S$ muons, and are proportional to
the asymmetry $\aslb$:
\begin{eqnarray}
a_S & = & c_b \aslb, \nonumber \\
A_S & = & C_b \aslb.
\label{as}
\end{eqnarray}
The dilution coefficients $c_b$ and $C_b$ are discussed in  Ref.~\cite {PRD}
and in Sec.~\ref{sec_Ab} below.

\begin{table}
\caption{\label{tab7}
Fractions of muon candidates in the inclusive muon sample
($f_\mu^i$) and in the like-sign dimuon sample ($F_\mu^i$, with two
entries per event).
}
\begin{ruledtabular}
\newcolumntype{A}{D{A}{\pm}{-1}}
\newcolumntype{B}{D{B}{-}{-1}}
\begin{tabular}{cBcc}
Bin & \multicolumn{1}{c}{Muon $p_T$ range (GeV)}       &  $f_\mu^i$ & $F_\mu^i$ \\
\hline
1  & 1.5\ B \ 2.5   & 0.0077 & 0.0774 \\
2  & 2.5\ B \ 4.2   & 0.2300 & 0.3227 \\
3  & 4.2\ B \ 5.6   & 0.4390 & 0.3074 \\
4  & 5.6\ B \ 7.0   & 0.1702 & 0.1419 \\
5  & 7.0\ B \ 10.0  & 0.1047 & 0.1057 \\
6  & 10.0\ B \ 25.0 & 0.0484 & 0.0449 \\
\end{tabular}
\end{ruledtabular}
\end{table}

Equations (\ref{inclusive_mu_a}) -- (\ref{as}) are used to measure the asymmetry $\aslb$.
The major contributions to the uncertainties on $\aslb$ are from the statistical uncertainty
on $A$ and the total uncertainty on $F^i_K$, $f^i_K$ and $\delta_i$.
To reduce the latter contributions,
we measure the asymmetry $\aslb$ using the asymmetry $A'$, which is defined as
\begin{equation}
A' \equiv A - \alpha a.
\label{aprime}
\end{equation}
Since the same physical processes contribute to both $F^i_K$ and $f^i_K$, their uncertainties
are strongly correlated, and therefore partially cancel in Eq.~(\ref{aprime}) for an appropriate choice
of the coefficient $\alpha$. The contribution from the asymmetry $\aslb$, however,
does not cancel in Eq.~(\ref{aprime}) because $c_b \ll C_b$ \cite{PRD}. Full details
of the measurements of different quantities entering in Eqs.~(\ref{inclusive_mu_a}) -- (\ref{as})
are given in Ref.~\cite{PRD}. The main improvements in the present analysis are related to muon
selection and the measurement of $F^i_K$ and $f^i_K$.
These modifications are described in Sections \ref{selection}, \ref{sec_fk} and \ref{sec_Fk}.

\section{Muon selection}
\label{selection}

\begin{figure}
\begin{center}
\includegraphics[width=0.45\textwidth]{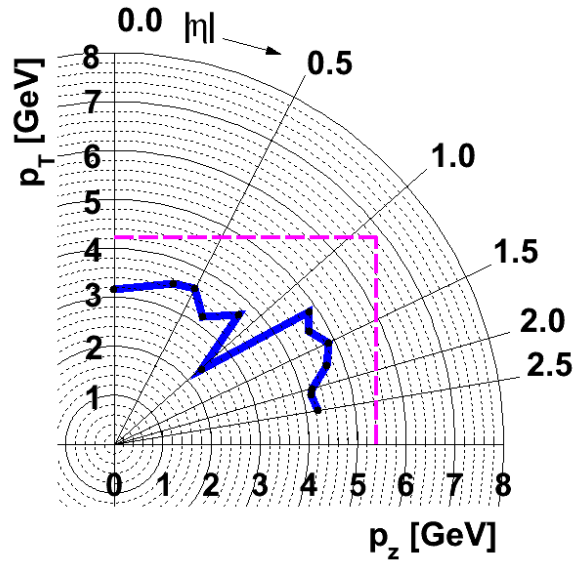}
\caption{(color online).
Smallest muon momentum required to penetrate the calorimeter and toroids at different
pseudorapidities, $|\eta|$ (solid line),
and the momentum selection used in this analysis (dashed line). 
}
\label{fig_minpt}
\end{center}
\end{figure}

The muon selection is similar to that
described in Ref.~\cite{PRD}.
The inclusive muon and like-sign dimuon samples are
obtained from data collected with single and dimuon triggers, respectively.
Charged particles with transverse momentum in the range $1.5 < p_T < 25$ GeV and with pseudorapidity
$|\eta| < 2.2$ \cite{rapidity} are considered as muon candidates.
The upper limit on $p_T$ is applied to suppress the
contribution of muons from $W$ and $Z$ boson decays. To
ensure that the muon candidate passes through the detector,
including all three layers of the muon system, we
require either $p_T > 4.2$ GeV or a longitudinal momentum
component $|p_z| > 5.4$ GeV. Muon candidates are selected
by matching central tracks with a segment reconstructed
in the muon system and by applying tight quality requirements
aimed at reducing false matching and background
from cosmic rays and beam halo. The transverse impact
parameter of the muon track relative to the reconstructed
$p \bar p$ interaction vertex must be smaller than 0.3 cm, with
the longitudinal distance from the point of closest approach
to this vertex smaller than 0.5 cm. Strict quality requirements
are also applied to the tracks and to
the reconstructed $p \bar p$ interaction vertex. The inclusive
muon sample contains all muons passing the selection requirements.
If an event contains more than one muon,
each muon is included in the inclusive muon sample. The
like-sign dimuon sample contains all events with at least
two muon candidates with the same charge. These two
muons are required to have an invariant mass greater
than 2.8 GeV to minimize the number of events in which
both muons originate from the same $b$ quark
(e.g., $b \rightarrow \mu$, $b \rightarrow c \rightarrow \mu$).
Compared to Ref.~\cite{PRD}, the following modifications to the muon
selection are applied:
\begin{itemize}
\item To reduce background from a mismatch of tracks in the
central detector with segments in the outer
muon system, we require that the sign of the
curvature of the track measured in the central tracker be the same as in the muon system.
This selection was not applied in Ref.~\cite{PRD}, and removes only about 1\% of the dimuon events.

\item To ensure that the muon candidate can penetrate all three layers of the muon detector,
we require either a transverse momentum $p_T > 4.2$ GeV, or
a longitudinal momentum component $|p_z| > 5.4$~GeV, instead of
$p_T > 4.2$ GeV or $|p_z| > 6.4$ GeV in Ref.~\cite{PRD}. With this change,
the number of like-sign dimuon events increases by 25\%,
without impacting the condition that the muon must penetrate
the calorimeter and toroids, as can be deduced from Fig.~\ref{fig_minpt}.

\item To reduce background from kaon and pion decays in flight, we require that the $\chi^2$
calculated from the difference
between the track parameters measured in the central tracker and in the muon system be
$\chi^2 < 12$ (for 4 d.o.f.) instead of 40 used in Ref.~\cite{PRD}. With this tighter selection,
the number of like-sign dimuon events is decreased by 12\%.

\end{itemize}

Compared to the selections applied in Ref.~\cite{PRD},
the total number of like-sign dimuon events after applying all these modifications is increased
by 13\% in addition to the increase due to the larger integrated luminosity
   of this analysis.

The muon charge is determined by the central tracker. The probability of charge mis-measurement
is obtained by comparing the charge measured by the central tracker and by the muon system and
is found to be less than 0.1\%.

The polarities of the toroidal and solenoidal magnetic fields are reversed
on average every two weeks so that the four solenoid-toroid polarity
combinations are exposed to approximately the same
integrated luminosity. This allows for a cancellation of first-order
effects related to the instrumental asymmetry~\cite{D01}.
To ensure such cancellation,
the events are weighted according to the number of events for each
data sample corresponding to a different configuration of the magnets' polarities.
These weights are given in Table~\ref{tab00}. During the data taking of
the last part of the sample, corresponding to
approximately 2.9~fb$^{-1}$ of $p \bar p$ collisions, the magnet polarities were
specially chosen to equalize the number of dimuon events with different polarities
in the entire sample. The weights in Table~\ref{tab00} are therefore closer to unity
compared to those used in Ref.~\cite{PRD}.

\begin{table}
\caption{\label{tab00}
Weights assigned to the events recorded with different solenoid and toroid polarities
in the inclusive muon and like-sign dimuon samples.
}
\begin{ruledtabular}
\newcolumntype{A}{D{A}{\pm}{-1}}
\newcolumntype{B}{D{B}{-}{-1}}
\begin{tabular}{cccc}
Solenoid & Toroid   & Weight & Weight \\
polarity & polarity & inclusive muon & like-sign dimuon \\
\hline
$-1$ & $-1$ & 0.994 & 0.964 \\
$-1$ & +1 & 1.000 & 1.000 \\
+1 & $-1$ & 0.985 & 0.958 \\
+1 & +1 & 0.989 & 0.978
\end{tabular}
\end{ruledtabular}
\end{table}

\section{Measurement of $\bm{f_K}$, $\bm{f_\pi}$, $\bm{f_p}$}
\label{sec_fk}

The fraction $f_K^i$ in the inclusive muon sample is measured using
$K^{*0} \to K^+ \pi^-$ decays, with the kaon identified as a muon
(see Ref.~\cite{PRD} for details). The transverse momentum of the $K^+$ meson
is required to be in the $p_T$ interval $i$. Since the momentum of a particle is
measured by the central tracking detector, a muon produced by a kaon is assigned
the momentum of this kaon
(a small correction for kaons decaying within the tracker volume is introduced later).
The fraction $f_{K^{*0}}^i$
of these decays is converted to the fraction $f_K^i$ using the relation
\begin{equation}
f_K^i = \frac{n_i(\ks)}{n_i(K^{*+} \to \ks \pi^+)} f_{K^{*0}}^i,
\label{fkst}
\end{equation}
where $n_i(\ks)$ and $n_i(K^{*+} \to \ks \pi^+)$ are the number of reconstructed
$\ks \to \pi^+ \pi^-$ and $K^{*+} \to \ks \pi^+$ decays, respectively. The transverse momentum
of the $\ks$ meson is required to be in the $p_T$ interval $i$. We
require in addition that one of the pions from the $\ks~\to~\pi^+\pi^-$ decay be identified
as a muon.
In the previous analysis \cite{PRD} the production of $K^{*+}$ mesons
was studied in a sample of events with an additional reconstructed muon,
but we did not require that this muon be associated with a pion
from $\ks \to \pi^+ \pi^-$ decay. The fraction of events containing $b$ and/or $c$ quarks
was therefore enhanced in the sample, which could result in a bias of the measured fraction
$f_K$. This bias does not exceed the systematic uncertainty of $f_K$ and its
impact on the $\aslb$ value is less than 0.03\%. The application of the new requirement
ensures that the flavor composition
in the selected $K^{*+} \to \ks \pi^+$ and $K^{*0} \to K^+ \pi^-$ samples
is the same and this bias is eliminated.

\begin{figure}
\begin{center}
\includegraphics[width=0.50\textwidth]{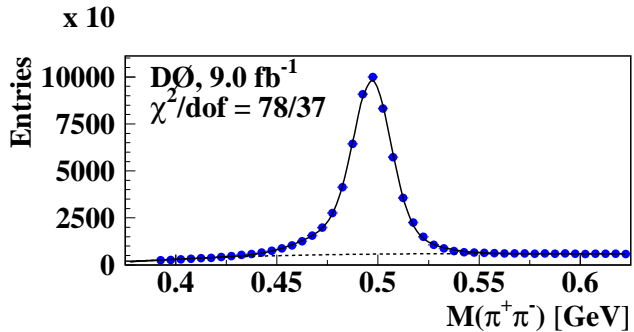}
\caption{(color online). The $\pi^+ \pi^-$ invariant mass distribution for $\ks$ candidates in the inclusive muon sample
with at least one pion identified as a muon with $4.2 < p_T(\ks) < 5.6$ GeV.
The solid line represents the result of the fit to the $\ks$ content,
and the dashed line represents the fitted background contribution. }
\label{fig-ks}
\end{center}
\end{figure}
\begin{figure}
\begin{center}
\includegraphics[width=0.50\textwidth]{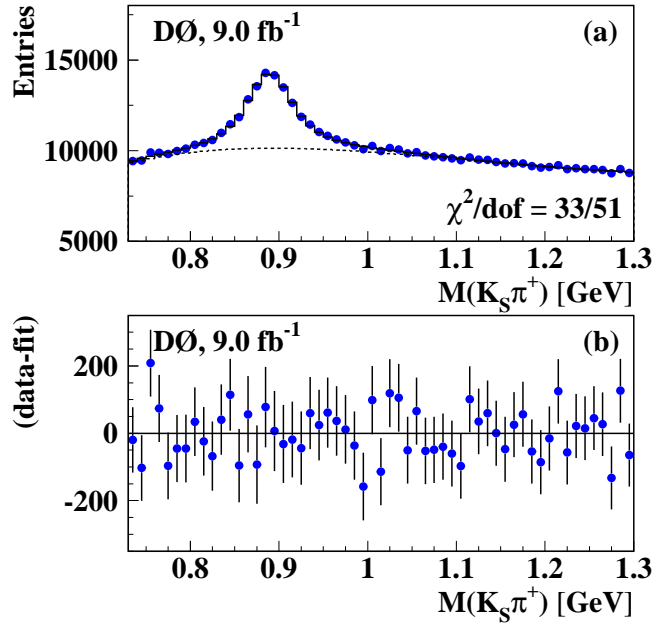}
\caption{(color online). (a) The $\ks \pi^+$ invariant mass distribution for $K^{*+}$ candidates
in the inclusive muon sample. The $\ks$ candidate is required to have
$480 < M(\pi^+ \pi^-) < 515$ MeV and
$4.2 < p_T(\ks) < 5.6$ GeV.
The solid line represents the result of the fit to the $K^{*+}$ content,
and the dashed line shows the background contribution.
(b) Difference between data and the result of the fit.
}
\label{fig-kstch}
\end{center}
\end{figure}
\begin{figure}
\begin{center}
\includegraphics[width=0.50\textwidth]{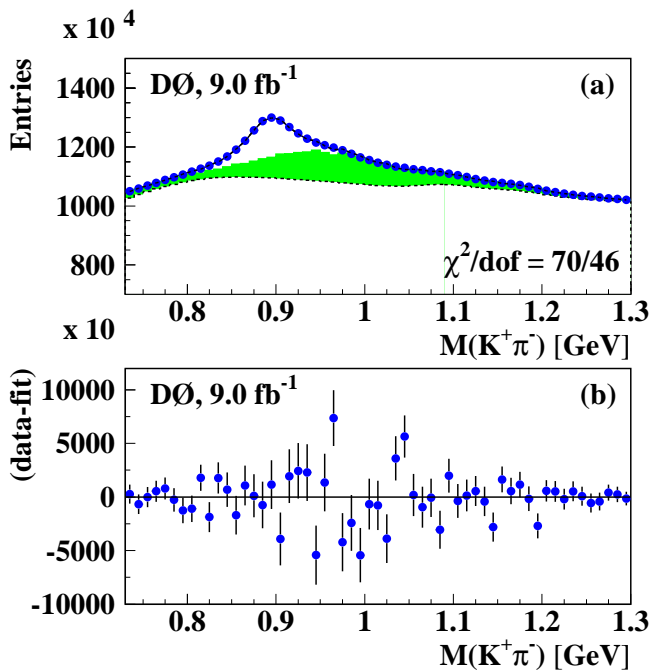}
\caption{(color online). (a) The $K^+ \pi^-$ invariant mass distribution for $K^{*0}$ candidates
in the inclusive muon sample for all kaons with $4.2 < p_T(K^+) < 5.6$ GeV.
The solid line corresponds to the result of the fit to the $K^{*0}$ content, and the
dashed line shows the contribution from combinatorial background. The shaded
histogram is the contribution from $\rho^0 \to \pi^+ \pi^-$ events. (b) Difference between data
and the result of the fit.}
\label{fig-kstn}
\end{center}
\end{figure}

The selection criteria and fitting procedures used to select and determine the
number of $\ks$, $K^{*+}$ and $K^{*0}$ events are given in Ref.~\cite{PRD}. As an example,
Fig.~\ref{fig-ks} displays the $\pi^+ \pi^-$ invariant mass distribution and the fitted
$\ks \to \pi^+ \pi^-$ candidates in the inclusive muon sample, with at least
one pion identified as a muon, for
$4.2 < p_T(\ks) < 5.6$ GeV. Figure~\ref{fig-kstch} shows the $\ks \pi^+$ mass distribution
and fit to $K^{*+} \to \ks \pi^+$ candidates for all $\ks$ candidates with
$4.2 < p_T(\ks) < 5.6$ GeV and $480 < M(\pi^+ \pi^-) < 515$ MeV.
Figure~\ref{fig-kstn} shows
the $K^+ \pi^-$ mass distribution and the fit result for $K^{*0} \to K^+ \pi^-$ candidates
for all kaons with $4.2 < p_T(K^+) < 5.6$ GeV. The $K^+ \pi^-$ mass distribution contains contributions
from light meson resonances decaying to $\pi^+ \pi^-$. The most important contribution comes from
the $\rho^0 \to \pi^+ \pi^-$ decay with $\pi \to \mu$. It produces a broad peak in the mass region close to
the $K^{*0}$ mass. The distortions in the background distribution due to other light resonances,
which are not identified explicitly, can also be seen in Fig.~\ref{fig-kstn}. Our background model
therefore includes the contribution of $\rho^0 \to \pi^+ \pi^-$ and two additional Gaussian terms to take into account
the distortions around 1.1 GeV. More details of the background description are given in Ref.~\cite{PRD}.

The measurement of the fractions $f_\pi$ and $f_p$ is also performed using the method
of Ref.~\cite{PRD}. The values of $f_K$ and $f_\pi$ are divided by the factors
$C_K$ and $C_\pi$, respectively, which take into account the fraction of kaons and pions
reconstructed by the tracking system before they decay. These factors are discussed in
Ref.~\cite{PRD}, and are determined through simulation.
Contrary to Ref.~\cite{PRD}, this analysis determines these factors separately for
kaons and pions. We find the values:
\begin{eqnarray}
C_K & = & 0.920 \pm 0.006, \nonumber \\
C_\pi & = & 0.932 \pm 0.006.
\end{eqnarray}
The uncertainties include contributions from the number
of simulated events and from the uncertainties in the momentum spectrum
of the generated particles.

\begin{figure}
\begin{center}
\includegraphics[width=0.50\textwidth]{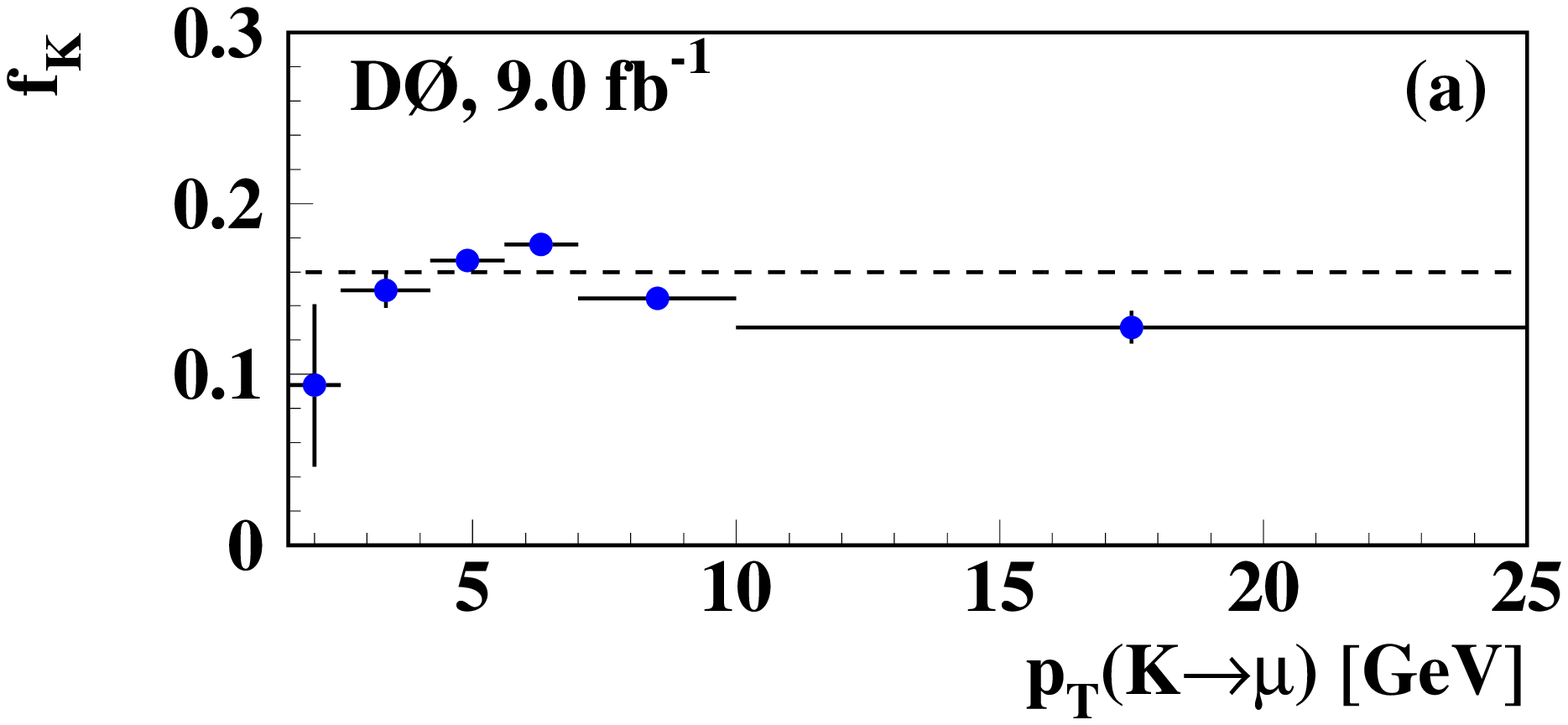}
\includegraphics[width=0.50\textwidth]{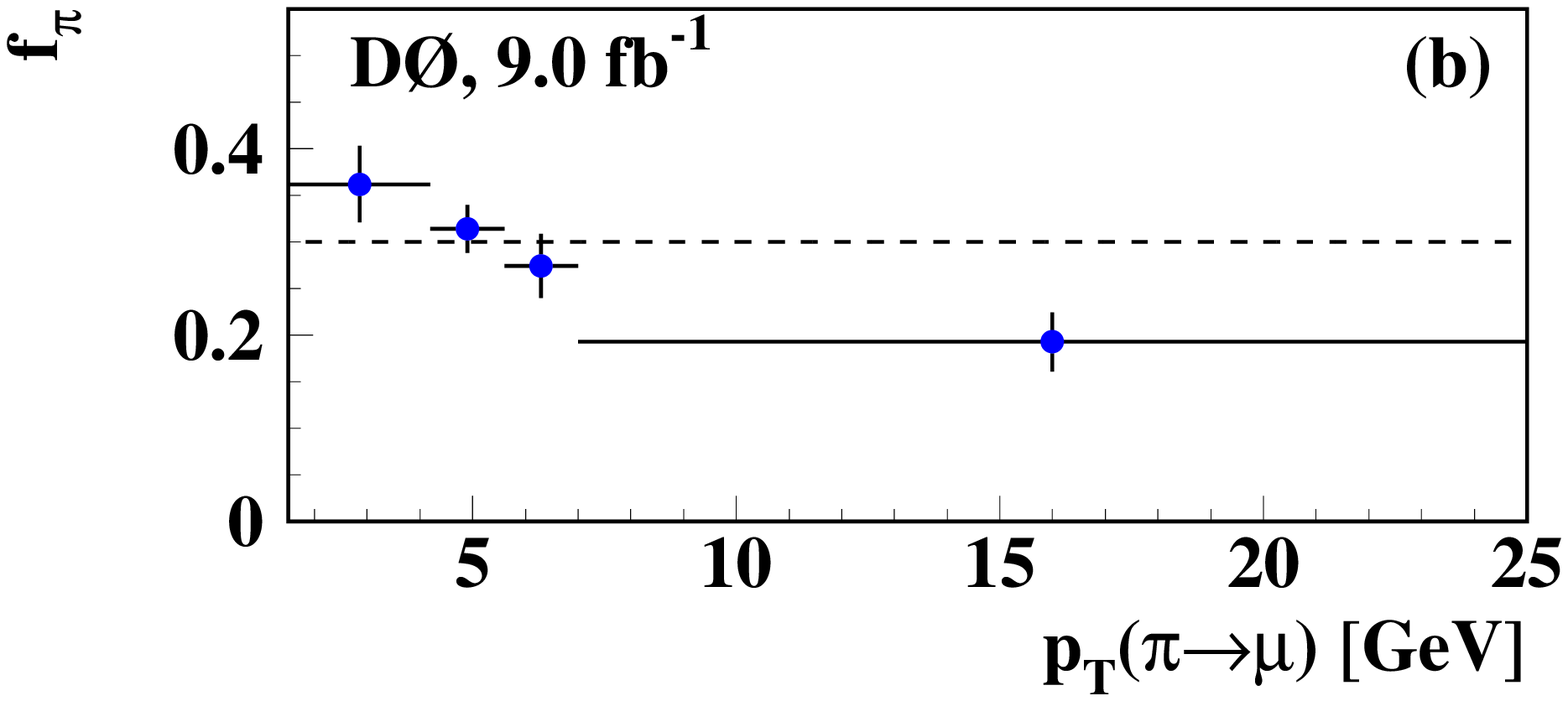}
\includegraphics[width=0.50\textwidth]{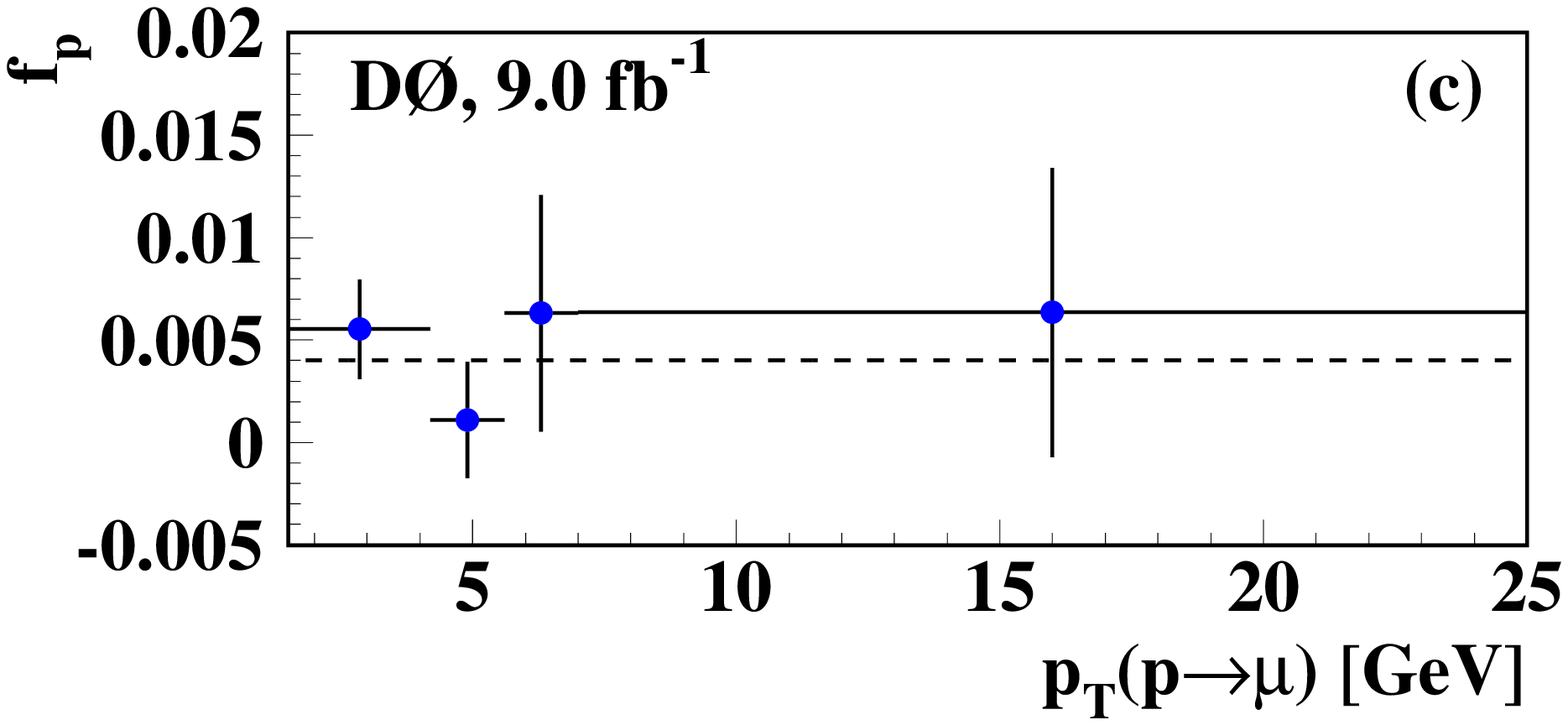}
\caption{(color online). The fraction of (a) $K \to \mu$ tracks, (b) $\pi \to \mu$ tracks and
(c) $p \to \mu$ tracks in the inclusive muon sample as a function of the
kaon, pion and proton $p_T$, respectively. The horizontal dashed lines
show the mean values.
}
\label{fig02}
\end{center}
\end{figure}

The values of $f_K$, $f_\pi$ and $f_p$ in different muon $p_T$ bins are
shown in Fig.~\ref{fig02} and in Table \ref{tab2}.
The changes in the muon candidates
 selection adopted here is the main source of differences
 relative to the corresponding values in Ref.~\cite{PRD}.
The fractions $f_\pi$ and $f_p$ are poorly
measured in bins 1 and 2, and bins 5 and 6 due to the small number of events, and their contents
are therefore combined through their weighted average.

%
\begin{table}
\caption{\label{tab2}
Fractions $f_K$, $f_\pi$, and $f_p$ for different $p_T$ bins.
The bottom row shows the weighted average
of these quantities obtained with weights given by the fraction
of muons in a given $p_T$ interval, $f^i_\mu$, in the inclusive muon
sample, see Table~\ref{tab7}. Only statistical uncertainties are given.
}
\begin{ruledtabular}
\newcolumntype{A}{D{A}{\pm}{-1}}
\newcolumntype{B}{D{B}{-}{-1}}
\begin{tabular}{cAcc}
Bin &  \multicolumn{1}{c}{$f_K \times 10^2$} & $f_\pi \times 10^2$ & $f_p \times 10^2$  \\
\hline
1     & 9.35  \ A \ 4.77   & \multirow{2}{*}{$36.20   \pm 4.12  $}
                            & \multirow{2}{*}{$0.55   \pm 0.24 $} \\
2     & 14.91  \ A \ 1.00   &                                       &  \\
3     & 16.65  \ A \ 0.41   & $31.42   \pm 2.57  $ & $ 0.11  \pm 0.29 $ \\
4     & 17.60  \ A \ 0.49   & $27.41   \pm 3.46  $ & $ 0.63  \pm 0.58 $ \\
5     & 14.43  \ A \ 0.45   & \multirow{2}{*}{$19.25   \pm 3.19  $}
                            & \multirow{2}{*}{$0.64    \pm 0.71 $}  \\
6     & 12.75  \ A \ 0.97   &                                       & \\ \hline
All & 15.96  \ A \ 0.24   & $30.01   \pm 1.60  $ & $ 0.38  \pm 0.17 $
\end{tabular}
\end{ruledtabular}
\end{table}

\section{Measurement of $\bm{F_K}$, $\bm{F_\pi}$, $\bm{F_p}$}
\label{sec_Fk}
The quantity $F_K$ is expressed as
\begin{equation}
F_K = R_K f_K,
\label{FK}
\end{equation}
where $R_K$ is the ratio of the fractions of muons produced by kaons in like-sign dimuon and
in inclusive muon data. For the $p_T$ interval $i$, $R_K$ is defined as
\begin{equation}
R_{K,i} = 2 \frac{N_i(K \to \mu)}{n_i(K \to \mu)}
\frac{n_i(\mu)}{N_i(\mu)},
\label{rk}
\end{equation}
where $N_i(K \to \mu)$ and $n_i(K \to \mu)$ are the number of
reconstructed $K$ mesons identified as muons in the like-sign dimuon and
in the inclusive muon samples, respectively. The transverse momentum of the $K$ meson is required to be
in the $p_T$ interval $i$. The quantities $N_i(\mu)$ and $n_i(\mu)$ are the number of muons
in the $p_T$ interval $i$. A multiplicative factor of two is included in Eq.~(\ref{rk})
because there are two muons in a like-sign dimuon event, and $F_K$ is normalized
to the number of like-sign dimuon events.

In the previous analysis \cite{PRD}, the quantity $F_K$
was obtained from a measurement of the $K^{*0}$ production rate.
Presenting it in the form of Eq.~(\ref{FK})
also allows the determination of
$F_K$ through an independent measurement of the fraction of $\ks$ mesons in
dimuon and in inclusive muon data where one of the pions from $\ks \to \pi^+ \pi^-$ decay
is identified as a muon. This measurement is discussed below. In addition,
Eq.~(\ref{FK}) offers an explicit separation of systematic uncertainties
associated with $F_K$.
The systematic uncertainty on the fraction $f_K$ affects the
two determinations of $\aslb$ based on Eqs.~(\ref{inclusive_mu_a}) and
(\ref{dimuon_A}) in a fully correlated way; therefore, its impact on the measurement
obtained using Eq.~(\ref{aprime}) is significantly reduced.
The systematic uncertainty on the ratio $R_K$ does not cancel in Eq.~(\ref{aprime}).
It is estimated directly from a comparison of
the values of $R_K$ obtained in two independent channels.

One way to measure $R_K$ is from the fraction of $K^{*0} \to K^+ \pi^-$ events in
the inclusive muon and like-sign dimuon data,
\begin{equation}
R_{K,i}(K^{*0}) = 2 \frac{N_i(K^{*0} \to \mu)}{n_i(K^{*0} \to \mu)}
\frac{n_i(\mu)}{N_i(\mu)},
\label{rkkst}
\end{equation}
where $N_i(K^{*0} \to \mu)$ and $n_i(K^{*0} \to \mu)$ are the number of
reconstructed $K^{*0} \to K^+ \pi^-$ decays, with the kaon identified as a muon in the like-sign
dimuon and in the inclusive muon samples, respectively. The transverse momentum of the $K$ meson
is required to be in the $p_T$ interval~$i$. The measurement using Eq.~(\ref{rkkst})
is based on the assumption
\begin{equation}
\frac{N_i(K^{*0} \to \mu)}{n_i(K^{*0} \to \mu)} = \frac{N_i(K \to \mu)}{n_i(K \to \mu)},
\end{equation}
which was validated through simulations in Ref.~\cite{PRD}.
The corresponding systematic uncertainty is discussed below.

\begin{figure}
\begin{center}
\includegraphics[width=0.50\textwidth]{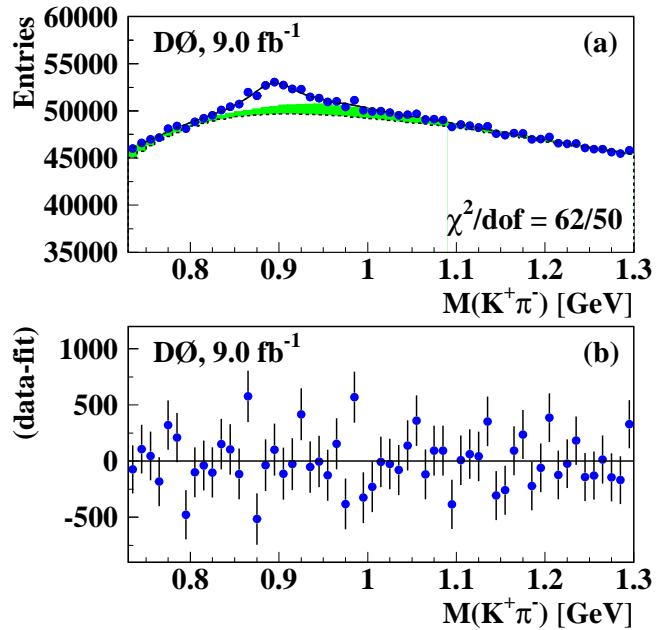}
\caption{(color online). (a) The $K^+ \pi^-$ invariant mass distribution of $K^{*0}$ candidates
in the like-sign dimuon sample for all kaons with $4.2 < p_T(K^+) < 5.6$ GeV.
The solid line corresponds to the result of the fit to the $K^{*0}$ content, and the
dashed line shows the contribution from combinatorial background. The shaded
histogram is the contribution from $\rho^0 \to \pi^+ \pi^-$ events. (b) Difference between data
and the result of the fit.}
\label{fig-kstn-2m}
\end{center}
\end{figure}

In Ref.~\cite{PRD}, the fractions $F_{K^{*0}}$ and $f_{K^{*0}}$ were obtained independently from a fit
of the $K^+ \pi^-$ invariant mass distribution in the like-sign dimuon
and inclusive muon sample, respectively.
Figure~\ref{fig-kstn-2m} shows
the same mass studies as in Fig.~\ref{fig-kstn}, but for
the like-sign dimuon sample.
The fit in both cases is complicated by the contribution from light meson resonances
that decay to $\pi^+ \pi^-$, producing a reflection in the $K^+ \pi^-$
invariant mass distribution. In addition, the detector resolution is not known a priori
and has to be included in the fit. All these complications are reduced significantly
or eliminated in the ``null-fit" method introduced in Ref.~\cite{PRD}, which is
used in this analysis to measure the ratio $R_K(K^{*0})$.

In this method, for each $p_T$ interval $i$, we define a set of distributions
$P_i(M_{K\pi}; \xi)$ that depend on a parameter~$\xi$:
\begin{equation}
P_i(M_{K\pi}; \xi) = N_i(M_{K\pi}) - \xi \frac{N_i(\mu)}{2 n_i(\mu)} n_i(M_{K \pi}),
\label{null_fit}
\end{equation}
where $N_i(M_{K\pi})$ and  $n_i(M_{K \pi})$ are the number of entries in the
$p_T$ bin $i$ of the $K^+ \pi^-$ invariant mass
distributions in the like-sign dimuon and inclusive muon samples, respectively.
For each value of $\xi$ the number of $K^{*0} \to K^+ \pi^-$ decays, ${\cal N}(K^{*0})$,
and its uncertainty, $\Delta {\cal N}(K^{*0})$, are measured from the $P_i(M_{K\pi}; \xi)$ distribution.
The value of $\xi$ for which ${\cal N}(K^{*0}) = 0$ defines
$R_{K,i}(K^{*0})$. The uncertainty $\sigma(R_{K,i})$ is determined from the condition that
${\cal N}(K^{*0}) = \pm \Delta {\cal N}(K^{*0})$ corresponding to $\xi = R_{K,i}(K^{*0}) \pm \sigma(R_{K,i})$.

\begin{figure}
\begin{center}
\includegraphics[width=0.50\textwidth]{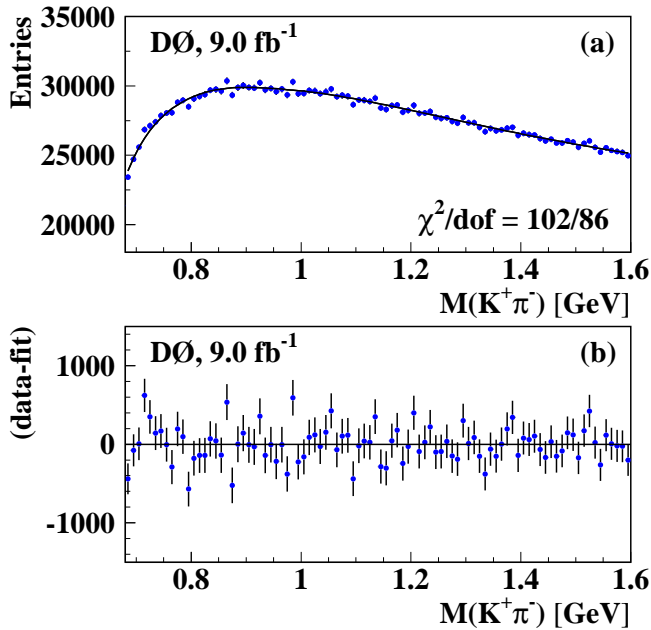}
\caption{(color online). (a) The $K^+ \pi^-$ invariant mass distribution $P_2(M_{K\pi}; \xi)$
obtained using Eq.~(\ref{null_fit}) for $\xi = 0.88$ for all kaons with
$4.2 < p_T(K^+) < 5.6$ GeV.
The dashed line shows the contribution from the combinatorial background.
(b) Difference between data and the result of the fit.}
\label{fig-null}
\end{center}
\end{figure}

The advantage of this method is that the influence of the detector resolution becomes minimal
for ${\cal N}(K^{*0})$ close to zero, and
the contribution from the peaking background
is reduced in $P_i(M_{K\pi}; \xi)$
to the same extent as the contribution of $K^{*0}$ mesons, and becomes
negligible when ${\cal N}(K^{*0})$ is close to zero.
As an example, Fig.~\ref{fig-null} shows the mass distribution $P_i(M_{K\pi}; \xi)$
for $\xi = 0.88$, for all kaons with $4.2 < p_T(K^+) < 5.6$ GeV. This distribution
is obtained from the distributions shown in Figs.~\ref{fig-kstn} and \ref{fig-kstn-2m},
using Eq.~(\ref{null_fit}). The contributions
of both $K^{*0} \to K^+ \pi^-$ and $\rho^0 \to \pi^+ \pi^-$, as well as any
other resonance in the background, disappear.
As a result, the fitting procedure becomes more robust, the fitting range
can be extended, and the resulting value of $R_K(K^{*0})$
becomes stable under a variation of the fitting parameters over a wider range.

The value of $R_K$ is also obtained from the production rate of $\ks$ mesons in the inclusive
muon and dimuon samples. We compute $R_{K,i}$ for a given $p_T$ interval $i$, as
\begin{equation}
R_{K,i}(\ks) = \frac{N_i(\ks \to \mu)}{n_i(\ks \to \mu)}
\frac{n_i(\mu)}{N_i(\mu)} \kappa_i,
\label{rks}
\end{equation}
where $N_i(\ks \to \mu)$ and $n_i(\ks \to \mu)$ are the number of
reconstructed $\ks \to \pi^+ \pi^-$ decays with one pion identified as a muon in the dimuon and
the inclusive muon data, respectively.
The correction factor $\kappa_i$ is discussed later in this section.
The measurement of $R_{K,i}$ using Eq.~(\ref{rks})
assumes isospin invariance and consequent equality of the ratio of production rates
in the dimuon and in the inclusive muon samples of $K^+$ and $\ks$ mesons, i.e.,
\begin{equation}
\frac{N_i(\ks \to \mu)}{n_i(\ks \to \mu)} = \frac{N_i(K \to \mu)}{n_i(K \to \mu)}.
\label{assume1}
\end{equation}
Since the charged kaon $p_T$ in Eq.~(\ref{assume1}) is required to be within the $p_T$ interval $i$,
the transverse momentum of the $\ks$ meson in Eq.~(\ref{rks})
is also required to be within the $p_T$ interval $i$.
We expect approximately the same number of
positive and negative pions from  $\ks \to \pi^+ \pi^-$ decays to be identified as a muon.
Therefore, we use both like-sign and opposite-sign dimuon
events to measure $N_i(\ks \to \mu)$ and we do not use the multiplicative factor of two
in Eq.~(\ref{rks}).  The requirement of having one pion identified
as a muon makes the flavor composition
in the samples of charged $K \to \mu$ events and $\ks \to \mu$
events similar.

\begin{figure}
\begin{center}
\includegraphics[width=0.50\textwidth]{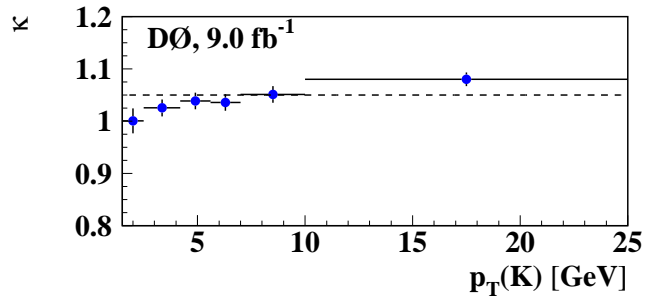}
\caption{(color online). The correction coefficient $\kappa$ as a function of the
kaon transverse momentum. The horizontal dashed line shows the mean value.
}
\label{fig04}
\end{center}
\end{figure}

The charges of the kaon and the additional muon in a dimuon event can be correlated, i.e.,
in general $N(K^+ \mu^+) \neq N(K^- \mu^+)$. However, the number of $N_i(\ks \to \mu)$ events
is not correlated with the charge of the additional muon, i.e.,
$N(\ks \to \mu^+, \mu^+) = N(\ks \to \mu^-, \mu^+)$. Since the ratio $R_{K,i}$ is
determined for the
sample of like-sign dimuon events, we apply in Eq.~(\ref{rks}) the correction factor $\kappa_i$,
defined as
\begin{equation}
\kappa_i \equiv \frac{2 (N(K^+ \mu^+) + c.c.)}
{(N(K^+ \mu^+) + N(K^- \mu^+) + c.c.)},
\label{kappa}
\end{equation}
to take into account the correlation between the charges of the kaon and muon.
The abbreviation ``c.c." in Eq.~(\ref{kappa}) denotes ``charge conjugate states".
The coefficients $\kappa_i$ are measured in data using the events with a reconstructed
$K^{*0} \to K^+ \pi^-$ decay and an additional muon. To reproduce the selection for the dimuon
sample \cite{PRD}, the invariant mass of the $K \mu$ system,
with the kaon assigned the mass of a muon,
is required to be greater than  2.8 GeV. The fitting procedure and selection criteria to measure
the number of $K^{*0}$ events are described in Ref.~\cite{PRD}. The values of $\kappa_i$
for different $p_T$ intervals are given in Fig.~\ref{fig04} and in Table~\ref{tab4}.

\begin{table}
\caption{\label{tab4}
Values of $\kappa$ in different $p_T$ bins. The bottom row shows their average.
Only statistical uncertainties are given.
}
\begin{ruledtabular}
\newcolumntype{A}{D{A}{\pm}{-1}}
\newcolumntype{B}{D{B}{-}{-1}}
\begin{tabular}{lA}
bin &  \multicolumn{1}{c}{$\kappa$} \\
\hline
1     & 1.005\ A \ 0.024 \\
2     & 1.025\ A \ 0.016 \\
3     & 1.038\ A \ 0.016 \\
4     & 1.036\ A \ 0.016 \\
5     & 1.051\ A \ 0.016 \\
6     & 1.080\ A \ 0.013 \\
\hline
Mean   & 1.046\ A \ 0.007 \\
\end{tabular}
\end{ruledtabular}
\end{table}

\begin{figure}
\begin{center}
\includegraphics[width=0.50\textwidth]{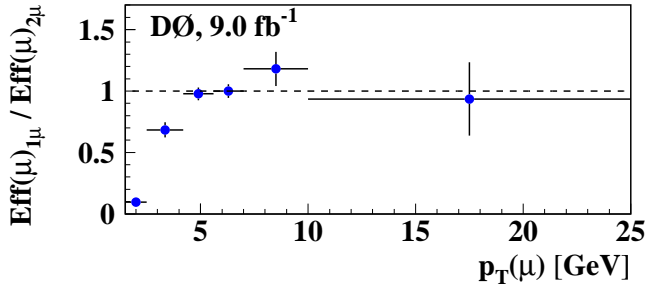}
\caption{(color online). The ratio of $\pi \to \mu$ detection efficiencies for the inclusive muon and
dimuon data as a function of the muon transverse momentum. The horizontal dashed line
shows the mean value for $p_T(K) > 4.2$ GeV.
}
\label{fig05}
\end{center}
\end{figure}

The average muon detection efficiency is different for the inclusive muon
and like-sign dimuon samples
because of different $p_T$ thresholds used in their triggers.
The difference in muon detection efficiency is large for muons with small $p_T$,
but it is insignificant for muons
above the inclusive-muon trigger threshold.
The ratio $N_i(\ks \to \mu)/n_i(\ks \to \mu)$ in Eq.~(\ref{rks})
is measured as a function of the transverse momenta of $\ks$ mesons, $p_T(\ks)$,
while the ratio $n_i(\mu)/N_i(\mu)$ is measured in bins of muon $p_T$.
Each $p_T(\ks)$ bin contains $\pi \to \mu$ with different $p_T(\pi \to \mu)$ values. The muon
detection efficiency therefore does not cancel in Eq.~(\ref{rks}),
and can affect the measurement
of $R_K(\ks)$. Figure~\ref{fig05} shows the ratio of $\pi \to \mu$ detection
efficiencies  in the inclusive muon and dimuon data.
To compute this ratio, we select the $\ks$ mesons in a given $p_T(\ks)$ interval.
The $p_T(\pi)$ distribution of pions produced in the $\ks \to \pi^+ \pi^-$ decay
with a given $p_T(\ks)$ is the same in the dimuon and inclusive muon data.
Therefore, any difference in this $p_T(\pi \to \mu)$ distribution
between dimuon and inclusive muon data is due to the $\pi \to \mu$ detection.
We compute the ratio of these $p_T(\pi \to \mu)$ distributions,
and normalize it such that it equals unity for $p_T(\pi \to \mu) > 5.6$ GeV.
The value of this $p_T$ threshold corresponds to the $p_T$ threshold for single muon
triggers.
Figure~\ref{fig05} presents the average of the ratios for different $p_T(\mu)$ intervals.
%
The ratio is suppressed for $p_T(\pi \to \mu) < 4.2$ GeV, and is consistent with a constant
for $p_T(\pi \to \mu) > 4.2$ GeV.
To remove the bias due to the trigger threshold, we measure $R_K(\ks)$
for events with $p_T(\pi \to \mu) > 4.2$ GeV. As a result, the ratio $R_K$ is not defined
for the first two $p_T$ bins in the $\ks$ channel.

\begin{figure}
\begin{center}
\includegraphics[width=0.50\textwidth]{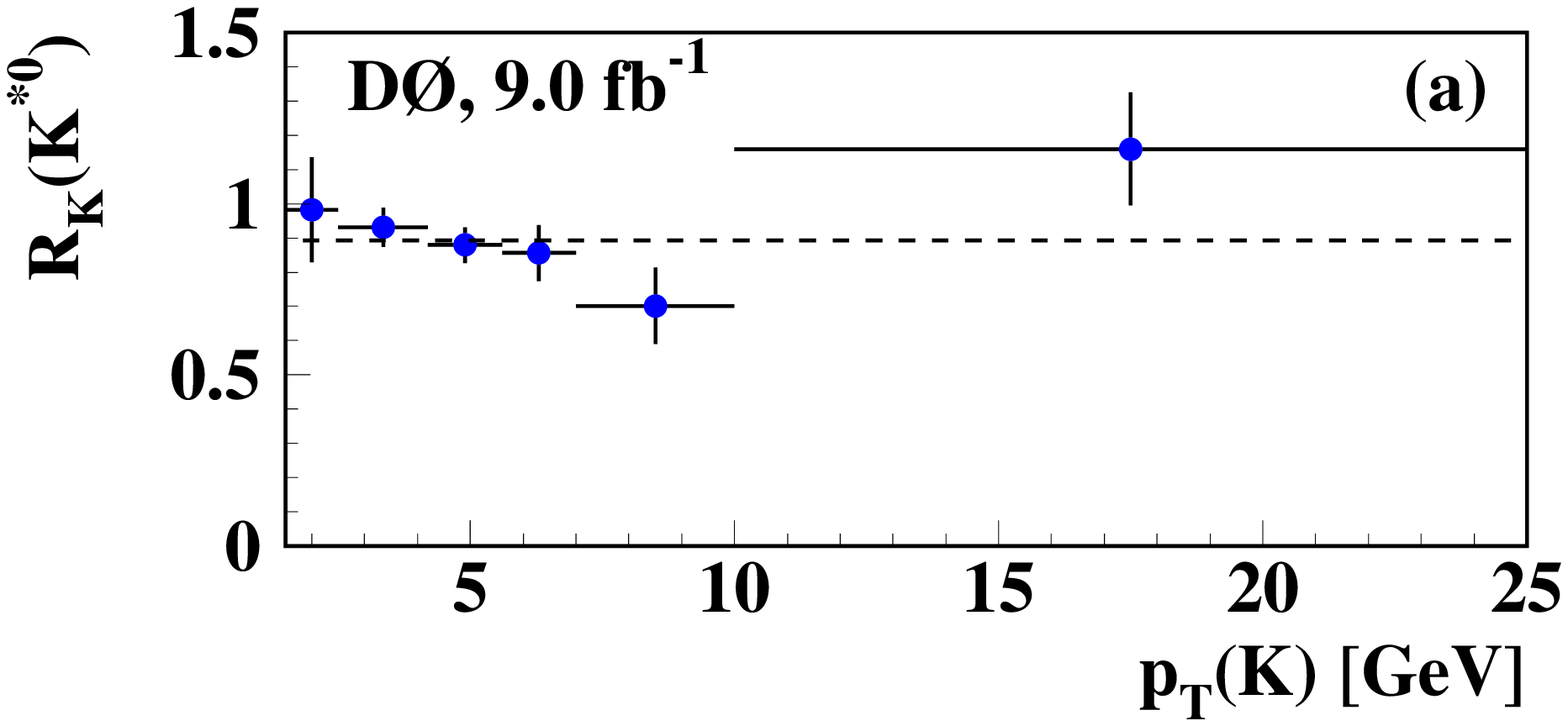}
\includegraphics[width=0.50\textwidth]{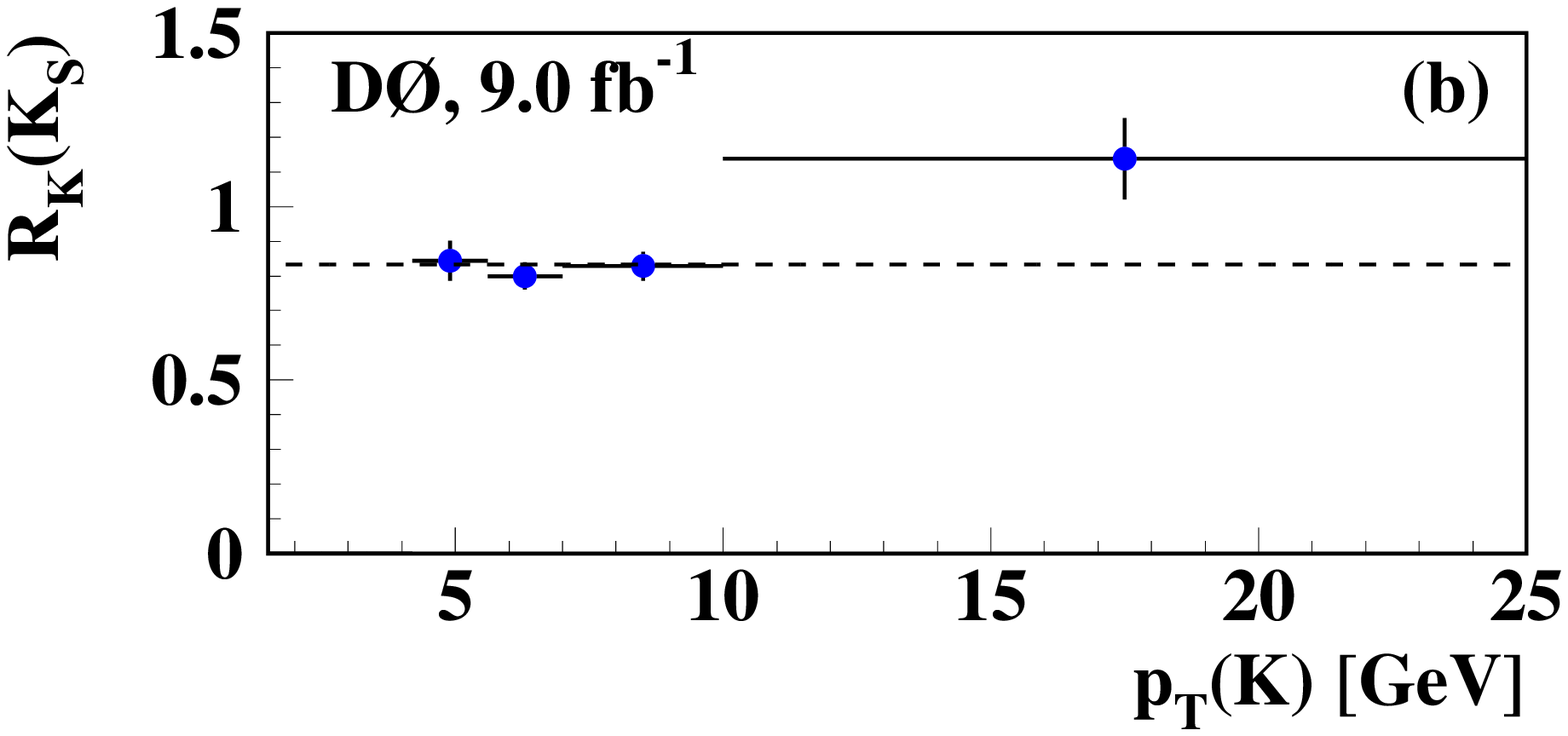}
\includegraphics[width=0.50\textwidth]{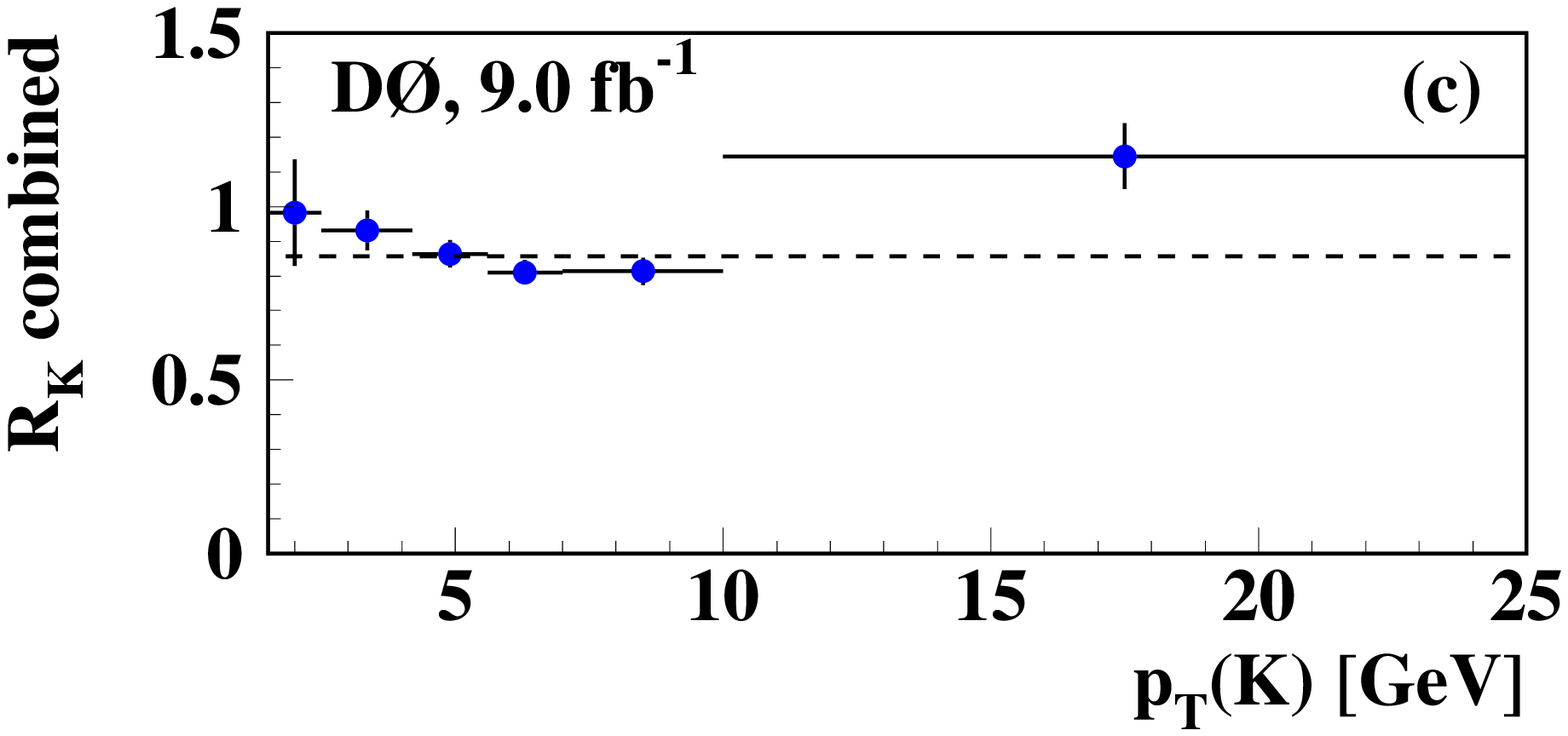}
\caption{(color online). The ratio $R_K$ obtained using (a) $K^{*0}$ production, (b) $\ks$ production, and
(c) combination of these two channels as a function of the
kaon transverse momentum. The horizontal dashed lines
show the mean values.
}
\label{fig03}
\end{center}
\end{figure}

\begin{figure}
\begin{center}
\includegraphics[width=0.50\textwidth]{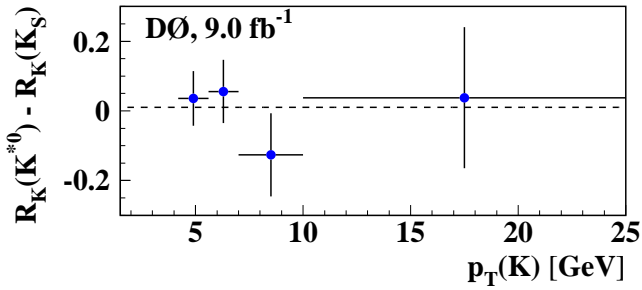}
\caption{(color online). The difference $R_K(K^{*0}) - R_K(\ks)$ as a function of kaon transverse momentum.
The horizontal dashed line shows the mean value.
}
\label{fig06}
\end{center}
\end{figure}

The values of $R_K(K^{*0})$ obtained through the null-fit method,
for different muon $p_T$ bins, are shown in Fig.~\ref{fig03}(a)
and in Table\ \ref{tab3}. The values of $R_K(\ks)$ are contained in Fig.~\ref{fig03}(b)
and in Table\ \ref{tab3}. The difference between the values of $R_K$ measured with $K^{*0}$ mesons
and with $\ks$ mesons is shown in Fig.~\ref{fig06}. The mean value of this difference is
\begin{equation}
\Delta R_K = 0.01 \pm 0.05,
\label{deltark}
\end{equation}
and the $\chi^2$/d.o.f. is 1.7/4.
We use two independent methods, each relying on different assumptions, to
measure the ratio $R_K$ and obtain results that are
   consistent with each other.
The methods are subject to different systematic uncertainties, and therefore provide an important
cross-check. As an independent cross-check,
the value of $R_K$ obtained in simulation is consistent with that measured in data,
see Sec.~\ref{sec_sim} for details. We take the average
of the two channels weighted by their uncertainties
as our final values of $R_K$ for $p_T(K) > 4.2$~GeV and use the values measured in
the $K^{*0}$ channel for $p_T(K) < 4.2$~GeV. These values are given in Table~\ref{tab3} and
in Fig.~\ref{fig03}(c). As we do not observe any difference between the two measurements,
we take half of the uncertainty of $\Delta R_K$ as the
systematic uncertainty of $R_K$. This corresponds to a relative uncertainty of 3.0\% on the
value of $R_K$. In our previous measurement \cite{PRD},
this uncertainty was 3.6\%, and was based on simulation of the events.

\begin{figure}
\begin{center}
\includegraphics[width=0.50\textwidth]{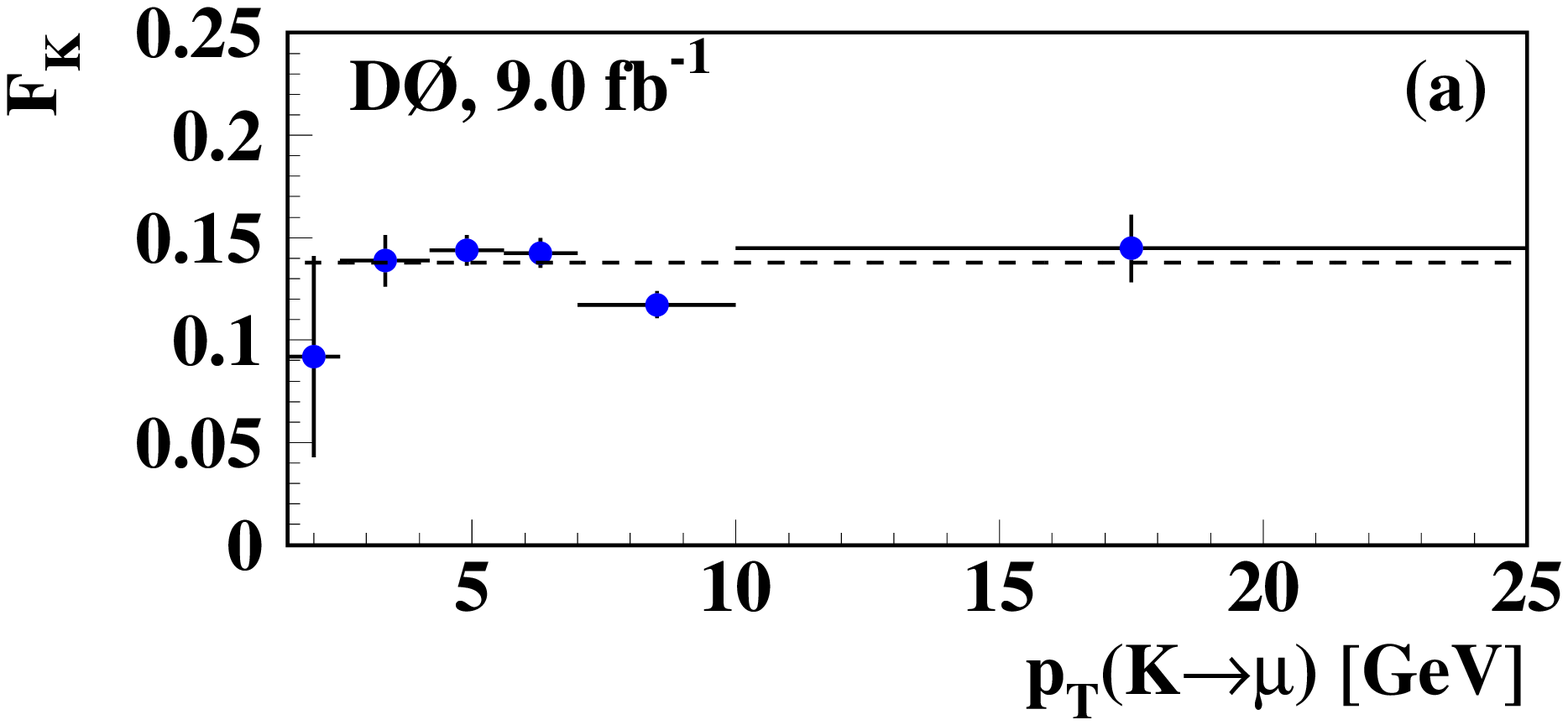}
\includegraphics[width=0.50\textwidth]{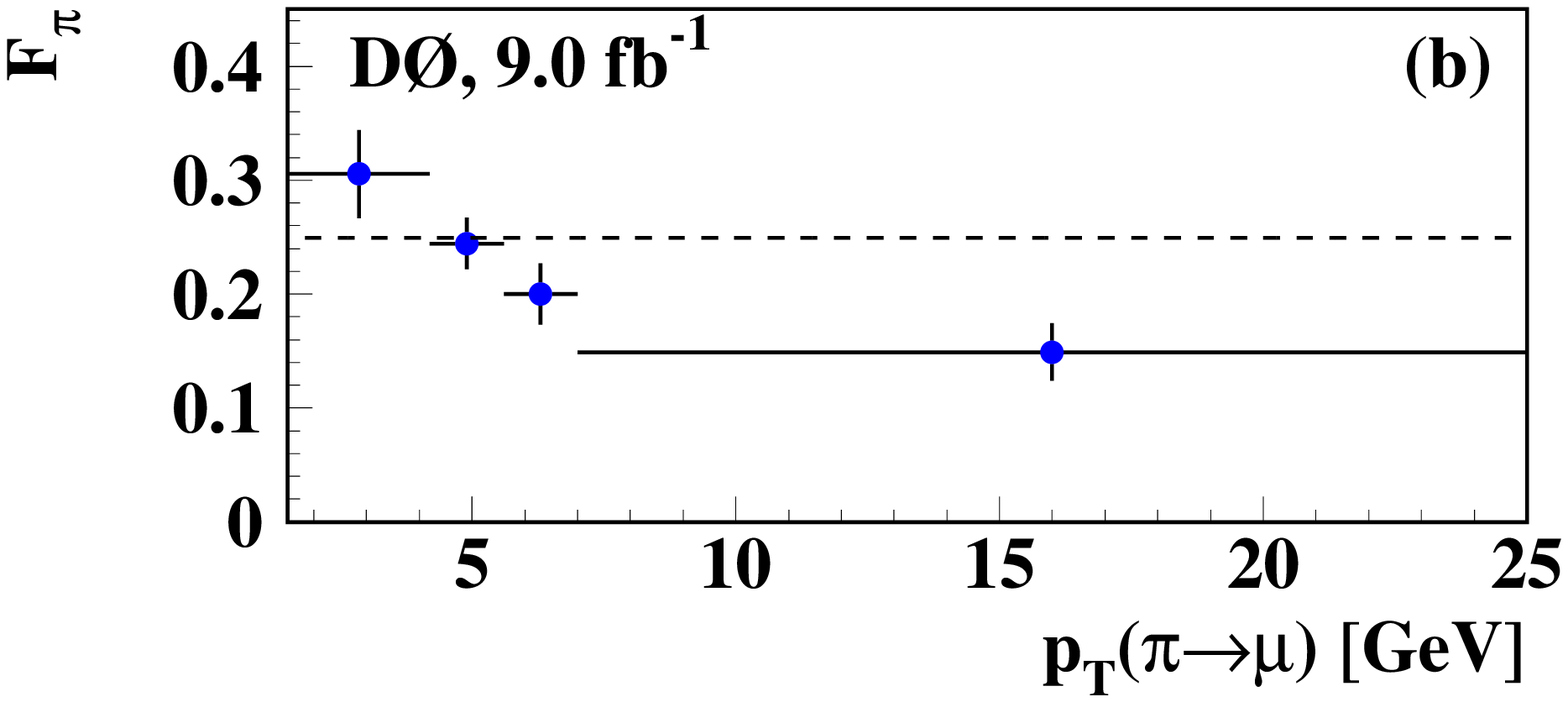}
\includegraphics[width=0.50\textwidth]{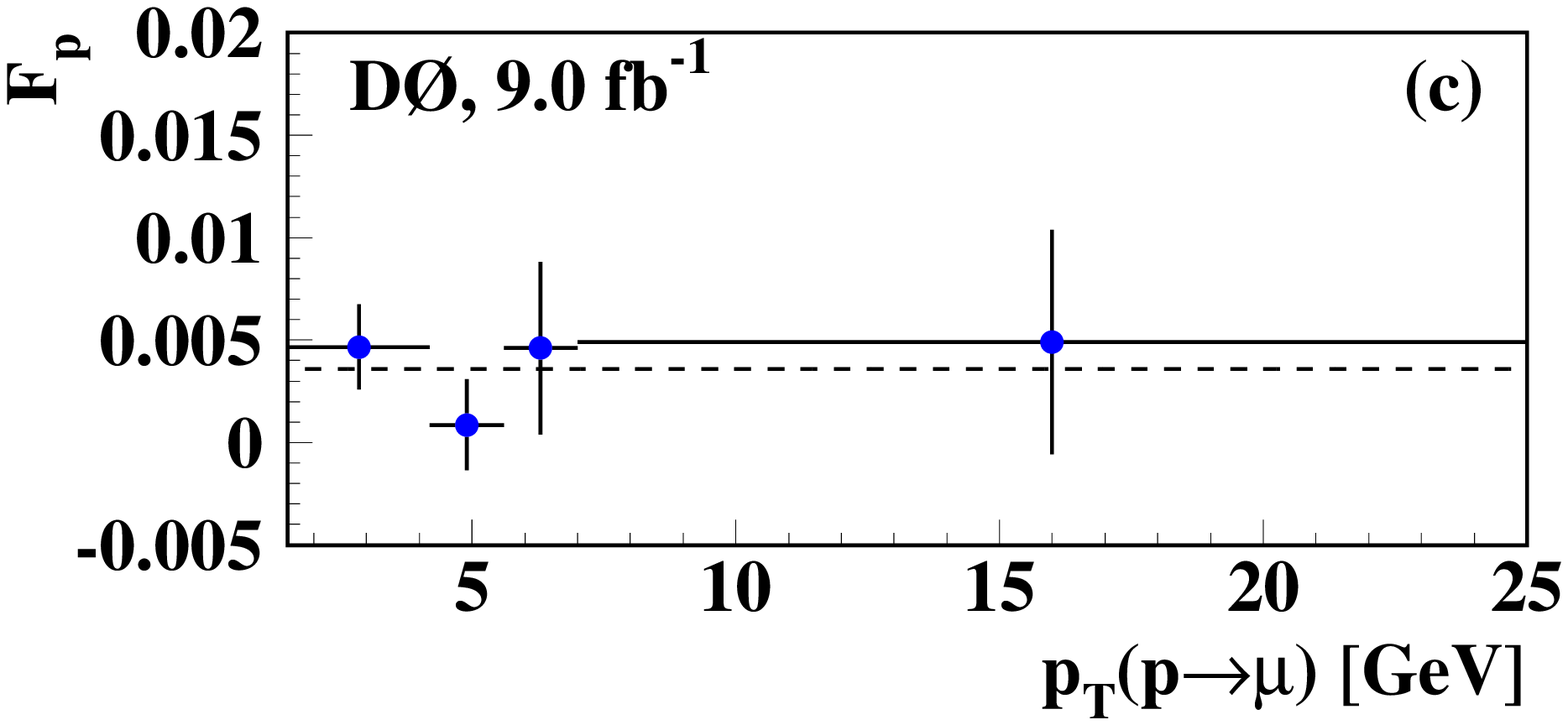}
\caption{(color online). The values of (a) $F_K$, (b) $F_\pi$ and
(c) $F_p$ in the like-sign dimuon sample as a function of
the kaon, pion and proton $p_T$, respectively.
The horizontal dashed lines show the mean values.
}
\label{fig07}
\end{center}
\end{figure}

Using the extracted values of $R_K$, we derive the values of $F_K$, $F_\pi$ and $F_p$.
The computation of $F_K$ is done using Eq.~(\ref{FK}), and we follow the procedure
described in Ref.~\cite{PRD} to determine $F_\pi$ and $F_p$. The results are shown in Fig.~\ref{fig07}
and in Table \ref{tab5}. The fractions $F_\pi$ and $F_p$ are poorly
determined for the lowest and highest $p_T$ because of the small number of events.
The content of bins
1 and 2, and bins 5 and 6 are therefore combined.

\begin{table}
\caption{\label{tab3}
Values of $R_K$ obtained using $K^{*0}$ and $\ks$ meson production in different $p_T$ bins.
The bottom row shows their average. Only statistical uncertainties are given. The ratio
$R_K$ in the $\ks$ channel is not measured in the first two bins, see Sec.\ \ref{sec_Fk}.
}
\begin{ruledtabular}
\newcolumntype{A}{D{A}{\pm}{-1}}
\newcolumntype{B}{D{B}{-}{-1}}
\begin{tabular}{lAAA}
bin &  \multicolumn{1}{c}{$R_K$ from $K^{*0}$} &
\multicolumn{1}{c}{$R_K$ from $\ks$} & \multicolumn{1}{c}{average $R_K$} \\
\hline
1     & 0.983\ A \ 0.154 &                  & 0.983\ A \ 0.154 \\
2     & 0.931\ A \ 0.058 &                  & 0.931\ A \ 0.058 \\
3     & 0.880\ A \ 0.052 & 0.844\ A \ 0.059 & 0.864\ A \ 0.039 \\
4     & 0.856\ A \ 0.082 & 0.800\ A \ 0.040 & 0.811\ A \ 0.036 \\
5     & 0.702\ A \ 0.112 & 0.828\ A \ 0.042 & 0.813\ A \ 0.039 \\
6     & 1.160\ A \ 0.165 & 1.138\ A \ 0.117 & 1.146\ A \ 0.095 \\
\hline
Mean   & 0.892\ A \ 0.032 & 0.834\ A \ 0.025 & 0.856\ A \ 0.020 \\
\end{tabular}
\end{ruledtabular}
\end{table}

\begin{table}
\caption{\label{tab5}
Values of $F_K$, $F_\pi$, and $F_p$ for different $p_T$ bins.
The last line shows the weighted average
of these quantities obtained with weights given by the fraction
of muons in a given $p_T$ interval $F^i_\mu$ in the dimuon sample, see Table \ref{tab7}.
Only statistical uncertainties are given.
}
\begin{ruledtabular}
\newcolumntype{A}{D{A}{\pm}{-1}}
\newcolumntype{B}{D{B}{-}{-1}}
\begin{tabular}{cAcc}
Bin &  \multicolumn{1}{c}{$F_K \times 10^2$} & $F_\pi \times 10^2$ & $F_p \times 10^2$  \\
\hline
1     & 9.19  \ A \ 4.90   & \multirow{2}{*}{$30.54   \pm 3.89  $}
                            & \multirow{2}{*}{$0.47   \pm 0.21 $} \\
2     & 13.88  \ A \ 1.26   &                                       &  \\
3     & 14.38  \ A \ 0.74   & $24.43   \pm 2.28  $ & $ 0.09  \pm 0.22 $ \\
4     & 14.26  \ A \ 0.74   & $19.99   \pm 2.67  $ & $ 0.46  \pm 0.42 $ \\
5     & 11.73  \ A \ 0.67   & \multirow{2}{*}{$14.90   \pm 2.55  $}
                            & \multirow{2}{*}{$0.49   \pm 0.55 $}  \\
6     & 14.48 \ A \ 1.64   &                                       & \\ \hline
All & 13.78  \ A \ 0.38   & $24.81   \pm 1.34  $ & $ 0.35  \pm 0.14 $
\end{tabular}
\end{ruledtabular}
\end{table}

\section{Systematic uncertainties for background fractions}
\label{sec_syst}

The systematic uncertainties for the background fractions are discussed in Ref.~\cite{PRD}, and we
only summarize the values used in this analysis. The systematic
uncertainty on the fraction $f_K$ is set to 9\% \cite{PRD}. The systematic
uncertainty on the ratio $R_K$, as indicated in Sec.~\ref{sec_Fk},
is set to half of the uncertainty on $\Delta R_K$ given in Eq.~(\ref{deltark}).
The systematic uncertainties on the
ratios of multiplicities $n_\pi/n_K$ and $n_p/n_K$ in $p \bar p$ interactions are set
to 4\% \cite{notation}.
These multiplicities are required to compute the quantities $f_\pi$, $f_p$.
The ratios $N_\pi/N_K$ and $N_p/N_K$, required to compute the quantities $F_\pi$ and $F_p$ \cite{PRD}
are assigned an additional 4\% systematic uncertainty.
The values of these uncertainties
are discussed in Ref.~\cite{PRD}.

\section{Measurement of $\bm{f_S}$, $\bm{F_{SS}}$}
\label{sec_fmu}

We determine the fraction $f_S$ of $S$ muons in the inclusive muon sample and
the fraction $F_{SS}$ of events with two $S$ muons in the like-sign dimuon sample
following the procedure described in Ref.~\cite{PRD}. We use the following value from
simulation
\begin{equation}
\frac{F_{LL}}{F_{SL}+F_{LL}} = 0.264 \pm 0.024,
\end{equation}
and obtain
\begin{eqnarray}
f_S & = & 0.536 \pm 0.017~({\rm stat}) \pm 0.043~({\rm syst}), \nonumber \\
F_{\rm bkg} & = & 0.389 \pm 0.019~({\rm stat}) \pm 0.038~({\rm syst}), \nonumber \\
F_{LL} & = & 0.082 \pm 0.005~({\rm stat}) \pm 0.010~({\rm syst}), \nonumber \\
F_{SL} & = & F_{\rm bkg} - 2 F_{LL}, \nonumber \\
F_{SS} & = & 0.692 \pm 0.015~({\rm stat}) \pm 0.030~({\rm syst}).
\end{eqnarray}
The difference between these values and that in Ref.~\cite{PRD} are due to
the increased statistics and the changes in the muon selection and in the analysis procedure.

\section{Measurement of $\bm{a_K}$, $\bm{a_\pi}$, $\bm{a_p}$, $\bm{\delta}$}
\label{sec_asym}
We measure all detector related asymmetries using the methods presented in Ref.~\cite{PRD}.
Muons from decays of charged kaons and pions and
from incomplete absorption of hadrons that penetrate
the calorimeter and reach the muon detectors (``punch-through"),
as well as false matches of central tracks to
segments reconstructed in the outer muon detector, are
considered as detector backgrounds. We use data to measure
the fraction of each source of background in both the
dimuon and inclusive muon samples, and the corresponding asymmetries.
Data are also used to determine the
intrinsic charge-detection asymmetry of the D0 detector.
Since the interaction length of the $K^+$ meson is greater
than that of the $K^-$ meson \cite{pdg}, kaons provide a positive
contribution to the asymmetries $A$ and $a$. The asymmetries
for other background sources (pions, protons and
falsely reconstructed tracks) are at least a factor of ten
smaller.

\begin{figure}
\begin{center}
\includegraphics[width=0.50\textwidth]{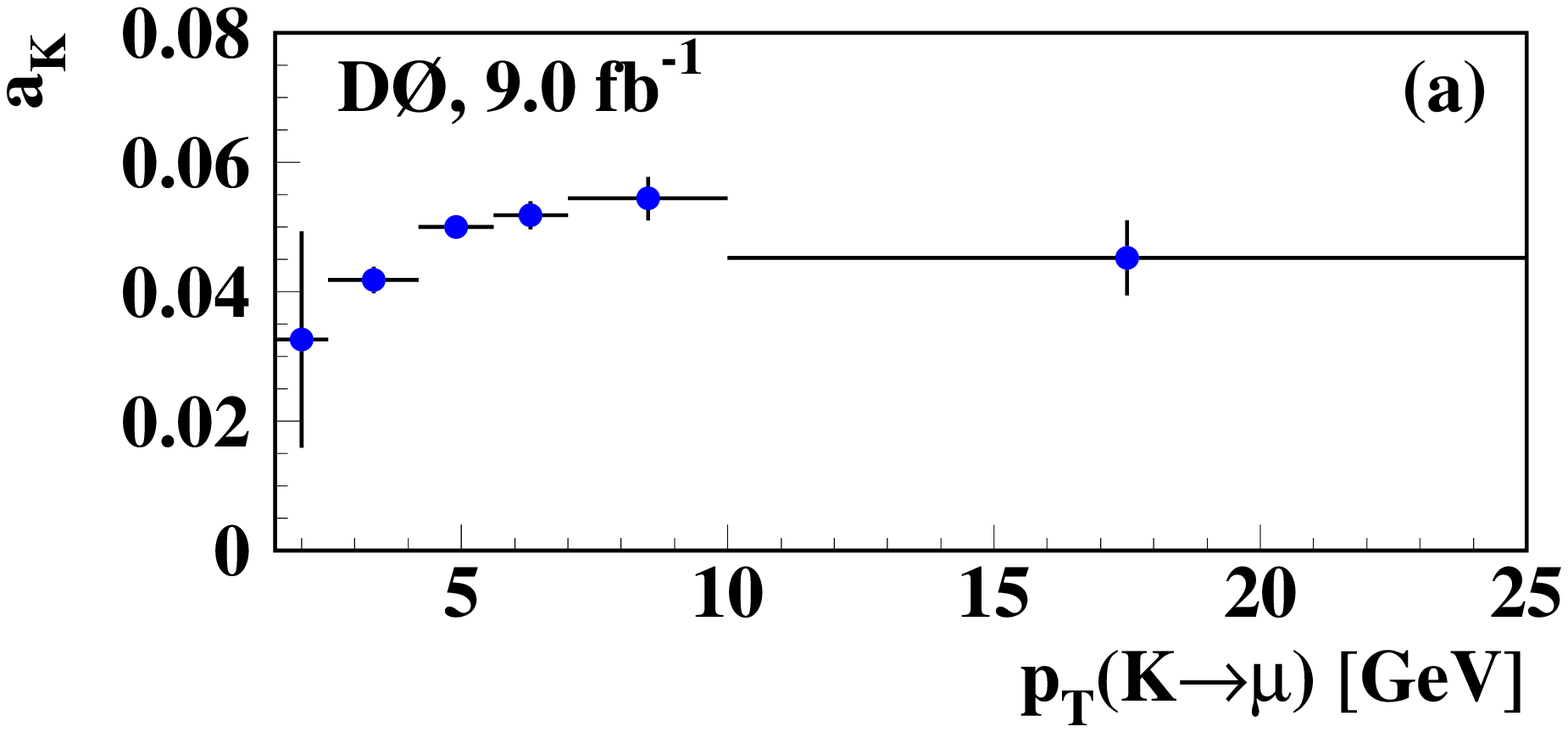}
\includegraphics[width=0.50\textwidth]{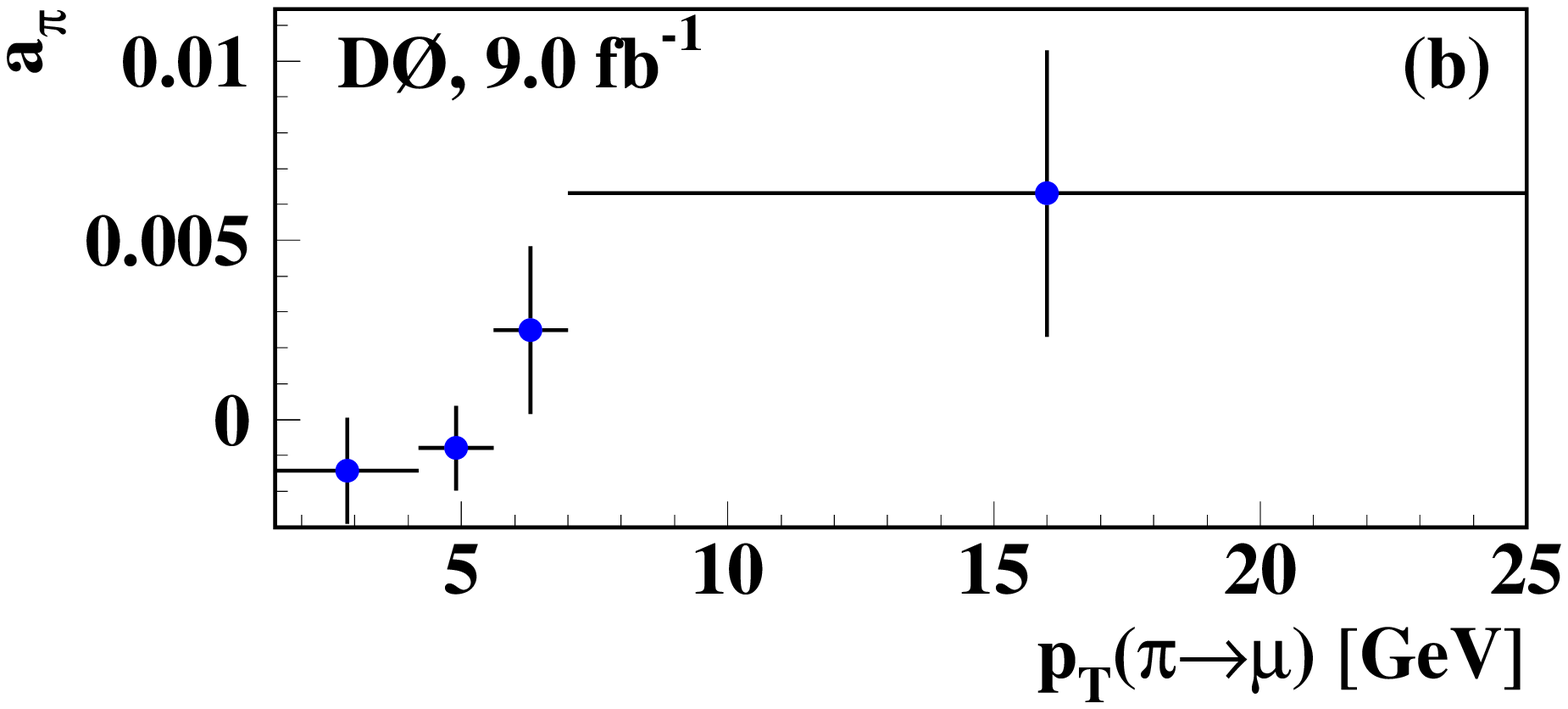}
\includegraphics[width=0.50\textwidth]{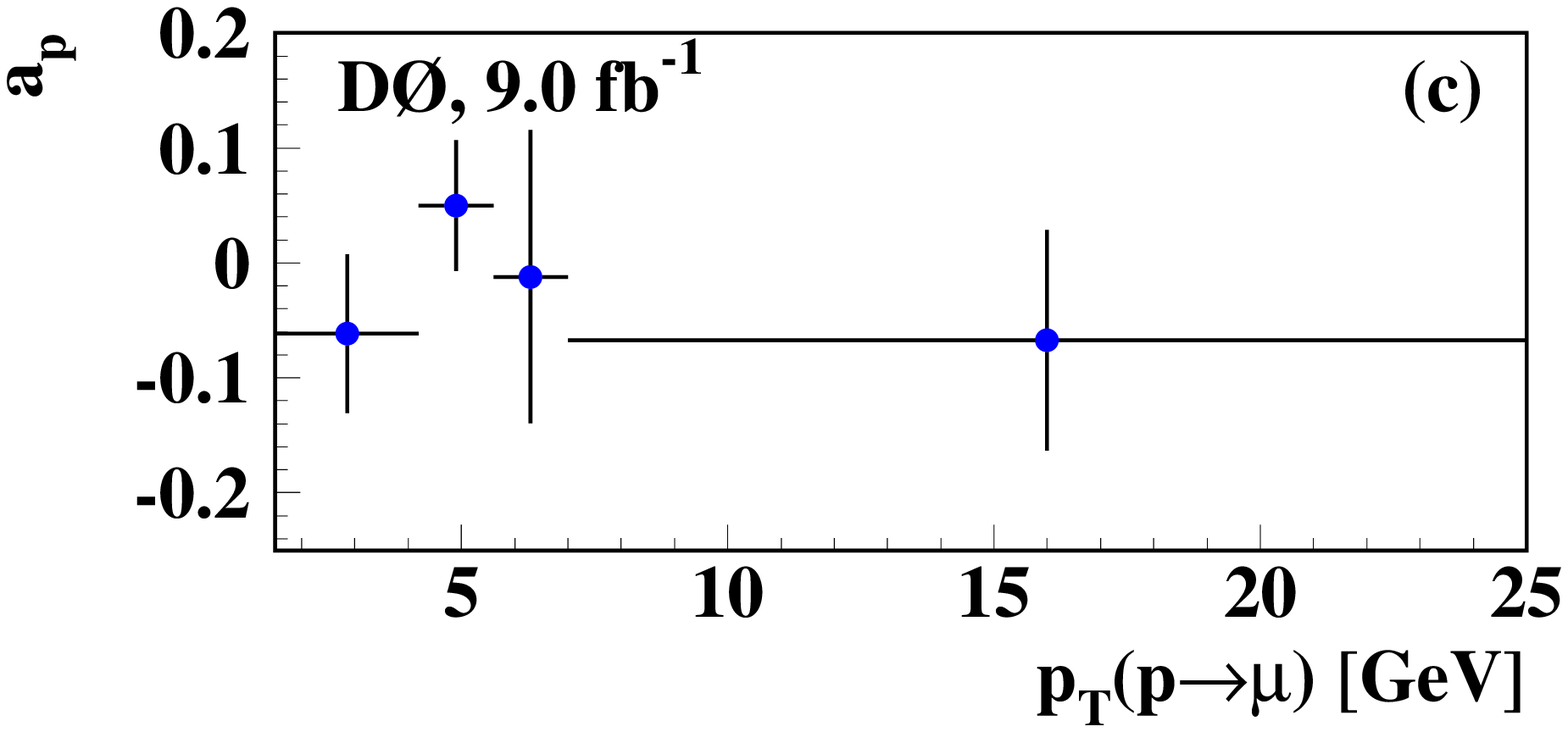}
\caption{(color online). The asymmetries (a) $a_K$, (b) $a_\pi$, and (c) $a_p$
as a function of the kaon, pion and proton $p_T$, respectively.}
\label{fig08}
\end{center}
\end{figure}

The results for the asymmetries $a_K$, $a_\pi$, and $a_p$ in different
muon $p_T$ bins are shown in Fig.~\ref{fig08} and Table \ref{tab6}.
The asymmetries $a_\pi$ and $a_p$ are poorly
measured in the first and last bins due to the small number of events. The content of bins
1 and 2, and bins 5 and 6 are therefore combined.

\begin{table}
\caption{\label{tab6}
Asymmetries $a_K$, $a_\pi$, and $a_p$ for different $p_T$ bins.
The bottom row shows the mean asymmetries averaged over the inclusive muon sample.
Only the statistical uncertainties are given.
}
\begin{ruledtabular}
\newcolumntype{A}{D{A}{\pm}{-1}}
\newcolumntype{B}{D{B}{-}{-1}}
\begin{tabular}{cAcc}
Bin &  \multicolumn{1}{c}{$a_K \times 10^2$} & $a_\pi \times 10^2$ & $a_p \times 10^2$  \\
\hline
1     & +3.26\ A \ 1.67 & \multirow{2}{*}{$-0.14 \pm 0.15$}
                            & \multirow{2}{*}{$-6.2 \pm 6.9$} \\
2     & +4.18\ A \ 0.20 &                                       &  \\
3     & +5.00\ A \ 0.13 & $-0.08 \pm 0.12$ & $+4.9 \pm 5.6$ \\
4     & +5.18\ A \ 0.22 & $+0.25 \pm 0.23$ & $-1.2 \pm 12.8$ \\
5     & +5.44\ A \ 0.34 & \multirow{2}{*}{$+0.63 \pm 0.40$}
                            & \multirow{2}{*}{$-6.8 \pm 9.6$}  \\
6     & +4.52\ A \ 0.57 &                                       & \\ \hline
All & +4.88\ A \ 0.09 & $-0.03 \pm 0.08$ & $-0.8 \pm 3.8$
\end{tabular}
\end{ruledtabular}
\end{table}

\begin{figure}
\begin{center}
\includegraphics[width=0.50\textwidth]{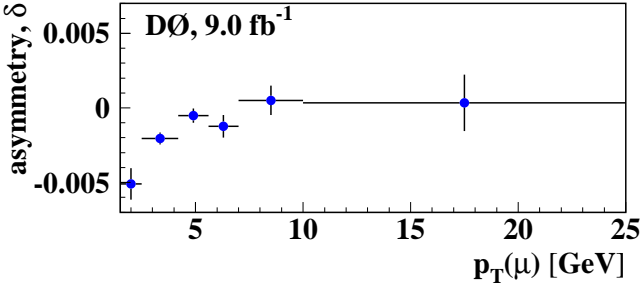}
\caption{(color online). Muon reconstruction asymmetry as a function of muon $p_T$.
}
\label{fig09}
\end{center}
\end{figure}

The small residual reconstruction asymmetry $\delta_i$ is measured
 using a sample of $J/\psi \to \mu^+\mu^-$ decays reconstructed
 from two central detector tracks, with at least one matching a
 track segment in the muon detector.
 The values of $\delta_i$ obtained as a function of muon $p_T$
are given in Table~\ref{tab8} and are shown in Fig.~\ref{fig09}.
The weighted averages for the residual muon asymmetry in the inclusive
muon and the like-sign dimuon samples, calculated using
weights given by the fraction of muons in each $p_T$ interval
$f_\mu^i$ ($F_\mu^i$) in the inclusive muon (dimuon) sample, are given by
\begin{eqnarray}
\label{delta}
\delta \equiv & \sum_{i=1}^6 f_\mu^i \delta_i = (-0.088 \pm 0.023) \%, \\
\Delta \equiv & \sum_{i=1}^6 F_\mu^i \delta_i = (-0.132 \pm 0.019) \%,
\label{Delta}
\end{eqnarray}
where only the statistical uncertainties are given.
The correlations among different $\delta_i$ are taken into account in the
uncertainties in Eqs.~(\ref{delta}) and~(\ref{Delta}).

\begin{table}
\caption{\label{tab8}
Muon reconstruction asymmetry $\delta_i$ for different muon $p_T$ bins.
Only the statistical uncertainties are given.
}
\begin{ruledtabular}
\newcolumntype{A}{D{A}{\pm}{-1}}
\newcolumntype{B}{D{B}{-}{-1}}
\begin{tabular}{cA}
Bin & \multicolumn{1}{c}{$\delta_i \times 10^2$} \\
\hline
1 &  -0.509\ A \ 0.106 \\
2 &  -0.205\ A \ 0.040 \\
3 &  -0.053\ A \ 0.048 \\
4 &  -0.124\ A \ 0.075 \\
5 &  +0.050\ A \ 0.099 \\
6 &  +0.034\ A \ 0.189
\end{tabular}
\end{ruledtabular}
\end{table}


\section{Corrections for background asymmetries}
\label{sec_sumback}
The corrections for the background and detector
 contributions to the measured raw asymmetries $a$ and $A$
 are obtained combining the results from 
Tables~\ref{tab7}, \ref{tab2}, \ref{tab5}, and \ref{tab6}, and summarized in
Tables~\ref{tab10} and~\ref{tab11}.
The values in the bottom row of these tables are computed by averaging the
corresponding quantities with weights given by the fraction of muons
in each $p_T$ interval $f_\mu^i$ ($F_\mu^i$) in the inclusive muon (dimuon) sample,
see Eqs.~(\ref{inclusive_mu_a}) and~(\ref{dimuon_A}). We use the mean values for
$f_\pi$, $F_\pi$, $f_p$, $F_p$, $a_\pi$, and $a_p$ in bins 1 and 2, and in bins 5 and 6,
as the number of events for those bins are not sufficient to perform
separate measurements.

\begin{table}
\caption{\label{tab10}
Corrections due to background asymmetries
$f_K a_K$, $f_\pi a_\pi$, and $f_p a_p$ for different $p_T$ bins.
The bottom row shows the weighted averages obtained using weights
given by the fraction of muons in a given $p_T$
interval, $f_\mu^i$, in the inclusive muon sample.
Only statistical uncertainties are given.
}
\begin{ruledtabular}
\newcolumntype{A}{D{A}{\pm}{-1}}
\newcolumntype{B}{D{B}{-}{-1}}
\begin{tabular}{cAcc}
Bin &  \multicolumn{1}{c}{$f_K a_K \times 10^2$} & $f_\pi a_\pi \times 10^2$ &
$f_p a_p \times 10^2$ \\
\hline
1     & +0.305\ A \ 0.220 & \multirow{2}{*}{$-0.052 \pm 0.054$}
                          & \multirow{2}{*}{$-0.034 \pm 0.041$}  \\
2     & +0.624\ A \ 0.052 &                                     &  \\
3     & +0.832\ A \ 0.030 & $-0.025 \pm 0.037$                  & $+0.005 \pm 0.016$ \\
4     & +0.912\ A \ 0.046 & $+0.068 \pm 0.065$                  & $-0.008 \pm 0.081$ \\
5     & +0.785\ A \ 0.054 & \multirow{2}{*}{$+0.121 \pm 0.079$}
                          & \multirow{2}{*}{$-0.043 \pm 0.077$} \\
6     & +0.577\ A \ 0.086 &                                     &  \\ \hline
All   & +0.776\ A \ 0.021 & $+0.007 \pm 0.027$                  & $-0.014 \pm 0.022$ \\

\end{tabular}
\end{ruledtabular}
\end{table}

\begin{table}
\caption{\label{tab11}
Corrections due to background asymmetries
$F_K a_K$, $F_\pi a_\pi$ and $F_p a_p$ for different $p_T$ bins.
The bottom row shows the weighted averages obtained using weights
given by the fraction of muons in a given $p_T$
interval, $F_\mu^i$, in the like-sign dimuon sample.
Only statistical uncertainties are given.
}
\begin{ruledtabular}
\newcolumntype{A}{D{A}{\pm}{-1}}
\newcolumntype{B}{D{B}{-}{-1}}
\begin{tabular}{cAcc}
Bin &  \multicolumn{1}{c}{$F_K a_K \times 10^2$} & \multicolumn{1}{c}{$F_\pi a_\pi \times 10^2$} &
\multicolumn{1}{c}{$F_p a_p \times 10^2$} \\
\hline
1     & +0.300\ A \ 0.222 & \multirow{2}{*}{$-0.044 \pm 0.046$}
                          & \multirow{2}{*}{$-0.029 \pm 0.035$}  \\
2     & +0.581\ A \ 0.060 &                                     &  \\
3     & +0.719\ A \ 0.042 & $-0.020 \pm 0.029$                  & $+0.004 \pm 0.012$ \\
4     & +0.739\ A \ 0.050 & $+0.050 \pm 0.047$                  & $-0.005 \pm 0.059$ \\
5     & +0.638\ A \ 0.054 & \multirow{2}{*}{$+0.094 \pm 0.062$}
                          & \multirow{2}{*}{$-0.033 \pm 0.060$} \\
6     & +0.655\ A \ 0.112 &                                     &  \\ \hline
All   & +0.633\ A \ 0.031 & $-0.002 \pm 0.023$                  & $-0.016 \pm 0.019$ \\
\end{tabular}
\end{ruledtabular}
\end{table}

\section{Coefficients $\bm{c_b}$ and $\bm{C_b}$}
\label{sec_Ab}

The dilution coefficients $c_b$ and $C_b$ in Eq.~(\ref{as}) are obtained through simulations using the method
described in Ref.~\cite{PRD}. Both coefficients depend on the value of the mean mixing probability, $\chi_0$.
We use the value obtained at LEP as averaged by HFAG \cite{hfag} for this measurement 
\begin{equation}
\chi_0({\rm HFAG}) = 0.1259 \pm 0.0042.
\end{equation}
To measure the weights for the different processes producing $S$ muons,
we correct the momentum distribution of generated
$b$ hadrons to match that in the data used in this analysis.
The determined weights \cite{notation} are given in Table \ref{tab22}.

\begin{table}
\caption{\label{tab22}
Heavy-quark decays contributing to the inclusive muon and like-sign dimuon
samples \cite{notation}.
The abbreviation ``non-osc" stands for ``non-oscillating," and ``osc" for ``oscillating."
All weights are computed using MC simulation.
}
\begin{ruledtabular}
\newcolumntype{A}{D{A}{\pm}{-1}}
\newcolumntype{B}{D{B}{-}{-1}}
\begin{tabular}{lll}
   & Process & Weight \\
\hline
$T_1$   & $b \to \mu^-X$ & $w_1 \equiv 1.$ \\
$T_{1a}$ & ~~~$b \to \mu^-X$ (non-osc) & $w_{1a} = (1-\chi_0) w_1$ \\
$T_{1b}$ & ~~~$\bar{b} \to b \to \mu^-X$ (osc) & $w_{1b} = \chi_0 w_1$ \\
$T_2$ &  $b \to c \to \mu^+X$ & $w_2 = 0.096 \pm 0.012$ \\
$T_{2a}$ & ~~~$b \to c \to \mu^+X$ (non-osc) & $w_{2a} = (1-\chi_0) w_2$ \\
$T_{2b}$ & ~~~$\bar{b} \to b \to c \to \mu^+X$ (osc) & $w_{2b} = \chi_0 w_2$ \\
$T_3$ & $b \to c \bar c q$ with $c \to \mu^+X$ or $\bar c \to \mu^-X$ & $w_3 = 0.064 \pm 0.006$ \\
$T_4$ & $\eta, \omega, \rho^0, \phi(1020), J/\psi, \psi' \to \mu^+ \mu^-$ & $w_4 = 0.021 \pm 0.002$ \\
$T_5$ & $b \bar b c \bar c$ with $c \to \mu^+X$ or $\bar c \to \mu^-X$ & $w_5 = 0.013 \pm 0.002$ \\
$T_6$ & $c \bar c$ with $c \to \mu^+X$ or $\bar c \to \mu^-X$ & $w_6 = 0.675 \pm 0.101$

\end{tabular}
\end{ruledtabular}
\end{table}

The uncertainty on the weights for the different processes contains contributions
from the uncertainty in the momentum of the generated $b$ hadrons and from the uncertainties
in $b$-hadron branching fractions. The difference in the weights with and without
the momentum correction 
contributes to the assigned uncertainties. Additional contributions
 to the uncertainties on the weights derive from the uncertainties on
 the inclusive branching fractions
$B \to \mu X$, $B \to c X$ and $B \to \bar{c} X$
\cite{pdg}.
We assign an additional uncertainty of 15\% to the weights $w_5$ and $w_6$ for
uncertainties on the cross sections for $c \bar{c}$ and $b \bar b c \bar c$ production.

The resulting $c_b$ and $C_b$ coefficients are found to be
\begin{eqnarray}
\label{cb}
c_b & = & +0.061 \pm 0.007, \\
C_b & = & +0.474 \pm 0.032.
\label{Cb}
\end{eqnarray}

\section{Asymmetry $\bm{\aslb}$}
\label{sec_ah}

The results obtained in Secs. \ref{sec_fk}--\ref{sec_Ab} are used to measure
the asymmetry $\aslb$ following the procedure of Ref.~\cite{PRD}.
Using $2.041 \times 10^9$ muons in the inclusive muon sample and $6.019 \times 10^6$ events in the
like-sign dimuon sample we obtain the following values for the uncorrected
asymmetries a and A:
\begin{eqnarray}
\label{value_a}
a & = & (+0.688 \pm 0.002) \%, \\
A & = & (+0.126 \pm 0.041) \%.
\label{value_A}
\end{eqnarray}
The difference between these values and those in Ref.~\cite{PRD} are due to
increased statistics and the changes in the muon selection.
The contributions from different background sources to the observed asymmetries $a$ and $A$
are summarized in Table~\ref{tab12}.
\begin{table}
\caption{\label{tab12}
Contribution of different background sources to the observed asymmetry in the inclusive muon and
like-sign dimuon samples. Only statistical uncertainties are given. }
\begin{ruledtabular}
\newcolumntype{A}{D{A}{\pm}{-1}}
\newcolumntype{B}{D{B}{-}{-1}}
\begin{tabular}{lcc}
Source & \multicolumn{1}{c}{inclusive muon} & \multicolumn{1}{c}{like-sign dimuon} \\
\hline
$(f_K a_K$ or $F_K A_K) \times 10^2$  &  $+0.776 \pm 0.021$ &  $+0.633 \pm 0.031$ \\
$(f_\pi a_\pi$ or
$F_\pi A_\pi) \times 10^2$            &  $+0.007 \pm 0.027$ &  $-0.002 \pm 0.023$ \\
$(f_p a_p$ or
$F_p A_p) \times 10^2$                &  $-0.014 \pm 0.022$ &  $-0.016 \pm 0.019$ \\
$[(1-f_{\rm bkg}) \delta$ or          &  \multirow{2}{*}{$-0.047 \pm 0.012$}
                                      &  \multirow{2}{*}{$-0.212 \pm 0.030$}     \\
$(2-F_{\rm bkg}) \Delta] \times 10^2$ &                     &                    \\ \hline
$(a_{\rm bkg}$ or $A_{\rm bkg}) \times 10^2$
                                      & $+0.722 \pm 0.042$  &  $+0.402 \pm 0.053$ \\ \hline
$(a$ or $A) \times 10^2$              &  $+0.688 \pm 0.002$ &  $+0.126 \pm 0.041$ \\
$[(a-a_{\rm bkg})$ or                 &  \multirow{2}{*}{$-0.034 \pm 0.042$}
                                      &  \multirow{2}{*}{$-0.276 \pm 0.067$}     \\
$(A-A_{\rm bkg})] \times 10^2$        &                     &
\end{tabular}
\end{ruledtabular}
\end{table}

\begin{figure}
\begin{center}
\includegraphics[width=0.50\textwidth]{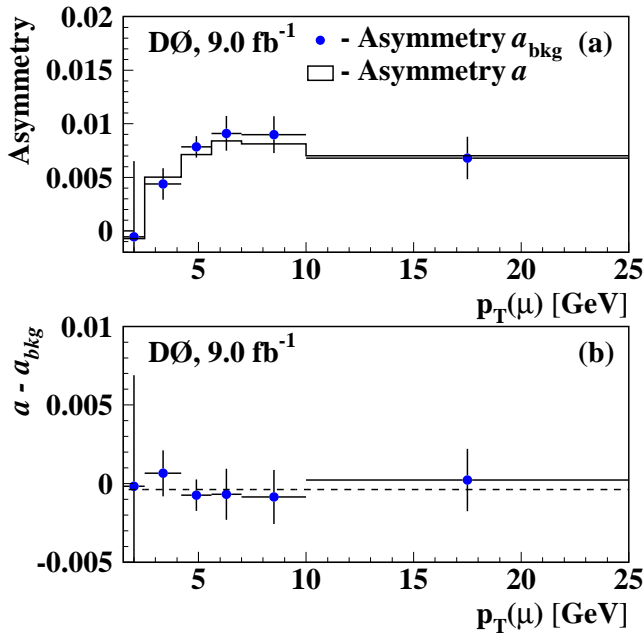}
\caption{(color online). (a)
The asymmetry $a_{\rm bkg}$ (points with error
bars representing the total uncertainties), expected from the measurements of the fractions and
asymmetries for background processes, is compared to the measured asymmetry $a$
for the inclusive muon sample (shown as a histogram, since the statistical uncertainties
are negligible).
The asymmetry from {\sl CP} violation is negligible
compared to the background in the inclusive muon sample.
(b)~The difference $a - a_{\rm bkg}$. The horizontal dashed line shows the mean value.
}
\label{fig10}
\end{center}
\end{figure}

The asymmetry $\aslb$, extracted from the asymmetry $a$ of the inclusive muon
sample using Eqs.~(\ref{inclusive_mu_a}) and~(\ref{cb}), is
\begin{equation}
\aslb = (-1.04 \pm 1.30~({\rm stat}) \pm 2.31~({\rm syst})) \%.
\label{ah1}
\end{equation}
The contributions to the uncertainty are given in Table~\ref{tab13}.
Figure~\ref{fig10}(a) shows a comparison of the asymmetry $a$ and
the background asymmetry, $a_{\rm bkg} = f_S \delta + f_K a_K + f_\pi a_\pi + f_p a_p$,
as a function of muon $p_T$. There is excellent agreement between these
two quantities, with $\chi^2/{\rm d.o.f.} = 0.8/6$ for their difference.
Figure~\ref{fig10}(b) shows the value of $f_S a_S = a - a_{\rm bkg}$,
which is consistent with zero. The values $a$ and $a_{\rm bkg}$ are given
in Table~\ref{tab14}. This result agrees with the expectation that
the value of the asymmetry $a$ is determined mainly by the
background, as the contribution from $\aslb$ is strongly
suppressed by the factor of $c_b = 0.061 \pm 0.007$.
The consistency of $\aslb$  with zero in Eq.~(\ref{ah1})
and the good description of the charge asymmetry $a$ for
different values of muon $p_T$ shown in Fig.~\ref{fig10}
constitute important tests of the validity of the background
model and of the method of analysis discussed in this Article.

\begin{table}
\caption{\label{tab13}
Sources of uncertainty on $\aslb$ from Eqs.~(\ref{ah1}),
(\ref{ah2}), and~(\ref{ah3}). The first nine rows contain statistical uncertainties,
while the next four rows reflect contributions from systematic uncertainties.
}
\begin{ruledtabular}
\newcolumntype{A}{D{A}{\pm}{-1}}
\newcolumntype{B}{D{B}{-}{-1}}
\begin{tabular}{cccc}
Source & $\delta(\aslb)\times 10^2$ & $\delta(\aslb)\times 10^2$ &
$\delta(\aslb) \times 10^2$ \\
       & Eq.~(\ref{ah1}) & Eq.~(\ref{ah2}) & Eq.~(\ref{ah3}) \\
\hline
$A$ or $a$ (stat)       & 0.068 & 0.121 & 0.132 \\
$f_K$ (stat)            & 0.472 & 0.064 & 0.028 \\
$R_K$ (stat)            &  N/A    & 0.059 & 0.065 \\
$P(\pitomu)/P(\ktomu)$  & 0.181 & 0.023 & 0.008 \\
$P(\ptomu)/P(\ktomu)$   & 0.323 & 0.026 & 0.002 \\
$A_K$                   & 0.458 & 0.052 & 0.037 \\
$A_\pi$                 & 0.802 & 0.067 & 0.030 \\
$A_p$                   & 0.584 & 0.050 & 0.020 \\
$\delta$ or $\Delta$    & 0.377 & 0.087 & 0.067 \\
\hline
$f_K$ (syst)            & 2.310 & 0.204 & 0.007 \\
$R_K$ (syst)            &   N/A   & 0.068 & 0.072 \\
$\pi$, $K$, $p$
multiplicity            & 0.067 & 0.019 & 0.017 \\
$c_b$ or $C_b$          & 0.121 & 0.052 & 0.056 \\
\hline
Total statistical       & 1.304 & 0.202 & 0.172 \\
Total systematic        & 2.313 & 0.222 & 0.093 \\
Total                   & 2.656 & 0.300 & 0.196
\end{tabular}
\end{ruledtabular}
\end{table}

\begin{table}
\caption{\label{tab14}
The measured asymmetry $a$ and the expected background asymmetry $a_{\rm bkg}$ in
the inclusive muon sample
for different $p_T$ bins. For the background asymmetry,
the first uncertainty is statistical, the second is systematic.
}
\begin{ruledtabular}
\newcolumntype{A}{D{A}{\pm}{-1}}
\newcolumntype{B}{D{B}{-}{-1}}
\begin{tabular}{cAA}
bin &  \multicolumn{1}{c}{$a \times 10^2$} &
\multicolumn{1}{c}{$a_{\rm bkg} \times 10^2$} \\
\hline
1     &-0.071\ A \ 0.025 &-0.055\ A \ 0.240 \pm 0.664 \\
2     &+0.503\ A \ 0.005 &+0.438\ A \ 0.089 \pm 0.117 \\
3     &+0.712\ A \ 0.003 &+0.785\ A \ 0.056 \pm 0.083 \\
4     &+0.841\ A \ 0.005 &+0.910\ A \ 0.124 \pm 0.105 \\
5     &+0.812\ A \ 0.007 &+0.897\ A \ 0.139 \pm 0.101 \\
6     &+0.702\ A \ 0.010 &+0.680\ A \ 0.189 \pm 0.059
\end{tabular}
\end{ruledtabular}
\end{table}

The second measurement of the asymmetry $\aslb$, obtained from the uncorrected asymmetry $A$ of
the like-sign dimuon sample using Eqs.~(\ref{dimuon_A}), (\ref{cb}) and~(\ref{Cb}), is
\begin{equation}
\aslb = (-0.808 \pm 0.202~({\rm stat}) \pm 0.222~({\rm syst})) \%,
\label{ah2}
\end{equation}
where we take into account that both $a_S$ and $A_S$ in Eq.~(\ref{dimuon_A}) are proportional
to $\aslb$, and that $F_{SS} C_b + F_{SL} c_b = 0.342 \pm 0.028$.
The contributions to the uncertainty of $\aslb$ for this measurement
are also listed in Table~\ref{tab13}.

The measurement of the asymmetry $\aslb$ using the linear combination given in Eq.~(\ref{aprime})
is performed following the procedure described in Ref.~\cite{PRD}.
We select the value of the
parameter $\alpha$ that minimizes the total uncertainty on the $\aslb$ measurement.
Appendix~\ref{app1} gives more details on this method of combination.
All uncertainties in Table \ref{tab13}, except the statistical
uncertainties on $a$, $A$, and $R_K$, are treated
as fully correlated. This leads to $\alpha = 0.89$,
and the corresponding value of the asymmetry $\aslb$ is
\begin{equation}
\label{ah3}
\aslb = (-0.787 \pm 0.172~({\rm stat}) \pm 0.093~({\rm syst}))\%.
\end{equation}
This value is used as the final result for \aslb.
It differs by 3.9 standard
deviations from the standard model prediction of $\aslb$ given in
Eq.~(\ref{in_aslbsm}).
The different contributions to the total uncertainty on $\aslb$ in Eq.~(\ref{ah3})
are listed in Table~\ref{tab13}.

The measured value of $\aslb$ places a constraint on the charge asymmetries
$\asld$ and $\asls$.
The asymmetry $\aslb$ is a linear combination of the semi-leptonic charge asymmetries from
$\Bd$ and $\Bs$ meson decays \cite{Grossman}.
The coefficients $C_d$ and $C_s$ in Eq.~(\ref{Ab_7}) depend
on the mean mixing probability and the production rate of $\Bd$ and $\Bs$ mesons.
We use the latest measurements of these quantities from LEP as averaged by HFAG \cite{hfag}
\begin{eqnarray}
\chi_0({\rm HFAG}) & = & 0.1259 \pm 0.0042, \\
f_d({\rm HFAG})    & = & 0.403  \pm 0.009, \\
f_s({\rm HFAG})    & = & 0.103  \pm 0.009,
\end{eqnarray}
and find the values given in Eq.~(\ref{Ab_8}).

\begin{figure}
\begin{center}
\includegraphics[width=0.50\textwidth]{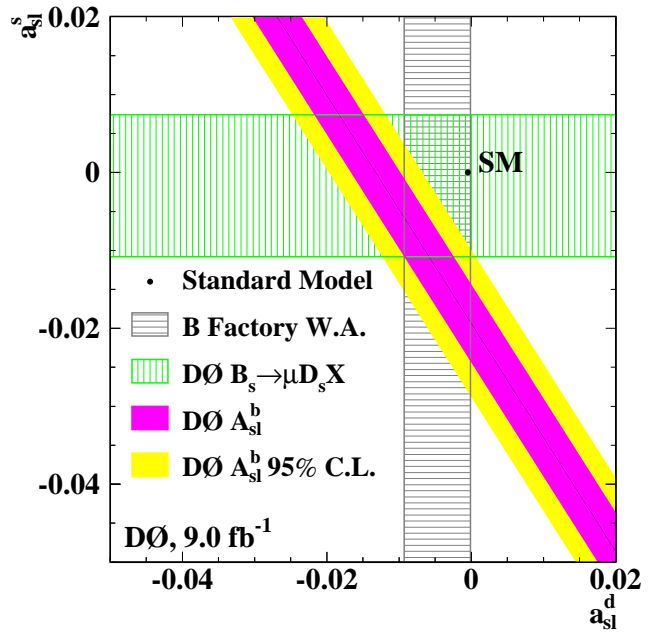}
\caption{(color online). Comparison of $\aslb$ in data
with the SM prediction for $\asld$ and $\asls$.
Also shown are the measurements of $\asld$~\cite{hfag} and
$\asls$~\cite{asl-d0}. The error bands represent the $\pm 1$ standard deviation uncertainties on
each individual measurement.
The 95\% C.L. band is also given for this $\aslb$ measurement.
}
\label{fig-ads1}
\end{center}
\end{figure}

Figure~\ref{fig-ads1} presents the measurement in the
($\asld,\asls$) plane together with the existing direct measurements of $\asld$ from the
$B$ factories \cite{hfag} and the independent D0 measurement of $\asls$ in
$\Bs \to \mu D_s X$ decays \cite{asl-d0}. All measurements are consistent.

The quantity $A_{\rm res}$ defined as
\begin{equation}
A_{\rm res} \equiv (A - \alpha a) - (A_{\rm bkg} - \alpha a_{\rm bkg})
\end{equation}
is the residual charge asymmetry of
like-sign dimuon events after subtracting all background contributions
from the raw charge asymmetry. This quantity does not depend on the interpretation
in terms of the charge asymmetry of semi-leptonic decays of $B$ mesons.
We obtain
\begin{equation}
\label{aslb-all}
A_{\rm res}  =  (-0.246 \pm 0.052~({\rm stat}) \pm 0.021~({\rm syst})) \%,
\end{equation}
The measured value of $A_{\rm res}$ differs by 4.2 standard deviations
from the standard model prediction
\begin{equation}
A_{\rm res}({\rm SM}) = (-0.009 \pm 0.002) \%.
\end{equation}

\section{Consistency checks}
\label{sec_consistency}

To study the stability of the result, we repeat this measurement with modified
selections, and with subsets of the available data. The only difference
compared to Ref.~\cite{PRD} is Test \textsf{D}, where we applied a stronger criterion on the muon
IP, following the suggestion of Ref.~\cite{Gronau-Rosner}. In all tests the
modified selections were applied to all muons. For completeness, we give
the full list of tests performed:
\begin{itemize}
\item{Test \textsf{A1}:} Using only the part of the data sample corresponding to the
first 2.8 fb$^{-1}$.
\item{Test \textsf{A2}:} Using only the part of the data sample corresponding to the
previous measurement with 6.1 fb$^{-1}$ \cite{PRD}.
\item{Test \textsf{A3}:} Using only the part of the data sample corresponding to the
last 2.9 fb$^{-1}$.
\item{Test \textsf{B}:} In addition to the reference muon selections \cite{PRD}, we require at least
three hits in the muon wire chambers (layers B or C), and lower the $\chi^2$
requirement for the fit to a track
segment reconstructed in  the muon detector.
\item{Test \textsf{C}:} Since background muons are mainly produced by decays of kaons
and pions, their track parameters measured in the central tracker and by the muon
system can differ. The background fraction therefore depends strongly on
the $\chi^2$ of the difference between these two measurements.
The requirement on this $\chi^2$ is changed from 12 to 4.
\item{Test \textsf{D}:} The maximum value of the IP
is changed from 0.3 to 0.012 cm. This test is also sensitive to
possible contamination from cosmic-ray muons.
\item{Test \textsf{E}:} Using low-luminosity data with fewer than three
interaction vertices.
\item{Test \textsf{F}:}
Using events corresponding to only two of four possible
configurations for the magnets, with identical solenoid and
toroid polarities.
\item{Test \textsf{G}:} Changing the minimum requirement on the invariant mass of the two muons
from 2.8 GeV to 12 GeV.
\item{Test \textsf{H}:} Using the same muon $p_T$ requirement, $p_T > 4.2$~GeV, for
the full detector acceptance.
\item{Test \textsf{I}:} Requiring the muon $p_T$ to be $p_T <7.0$~GeV.
\item{Test \textsf{J}:} Requiring the azimuthal angle $\phi$ of the muon track to be in the range
$0 < \phi < 4$ or $5.7 < \phi < 2 \pi$. This selection excludes muons with
reduced muon identification efficiency in the region of the support structure of the detector.
\item{Test \textsf{K}:} Requiring the muon $\eta$ to be in the range $ |\eta| < 1.6$. This test
is sensitive to possible contamination from muons associated
with beam halos.
\item{Test \textsf{L}:} Requiring the muon $\eta$ to have $ |\eta| < 1.2$
or $1.6 < |\eta| < 2.2$.
\item{Test \textsf{M}:} Requiring the muon $\eta$ to be in the range $ |\eta| < 0.7$
or $1.2 < |\eta| < 2.2$.
\item{Test \textsf{N}:} Requiring the muon $\eta$ to be in the range
$0.7 < |\eta| < 2.2$.
\item{Test \textsf{O}:} Using like-sign dimuon events that pass at least one single muon trigger,
while ignoring the requirement for a dimuon trigger.
\item{Test \textsf{P}:} Using like-sign dimuon events passing both single muon and dimuon triggers.
\end{itemize}

A summary of these studies is presented in Tables~\ref{tab20} and~\ref{tab21}.
The last row, denoted as ``Significance",
gives the absolute value of the difference between the reference result (column \textsf{Ref})
and each modification, divided by its uncertainty, taking
into account the overlap in events between the reference and test samples.
Both statistical and systematic uncertainties are used in the calculation of the
significance of the difference. The $\chi^2$ of these tests
defined as the sum of the square of all significances is $\chi^2 = 17.1$ for 18 tests.
These tests demonstrate the stability of the measured asymmetry $\aslb$,
and provide a confirmation of the validity of the method.

\begin{table*}
\caption{\label{tab20}
Measured asymmetry $\aslb$ for the reference selection (column \textsf{Ref}) and for
samples used in Tests \textsf{A -- H}.
}
\begin{ruledtabular}
\newcolumntype{A}{D{A}{\pm}{-1}}
\newcolumntype{B}{D{B}{.}{-1}}
\begin{tabular}{lrrrrrrrrrrr}
& \textsf{Ref} & \textsf{A1} & \textsf{A2} & \textsf{A3} & \textsf{B} & \textsf{C} & \textsf{D} &
\textsf{E} & \textsf{F} & \textsf{G} & \textsf{H}
\\ \hline
$N(\mu \mu) \times 10^{-6}$  &  6.019   &  1.932  & 3.991 &  2.028 & 4.466 &  3.280 &  2.857 &  3.128 &  3.012 &  2.583 &  2.220 \\
$a \times 10^2$              & +0.688   & +0.703  & +0.680 & +0.702 & +0.548 & +0.325 & +0.835 & +0.682 & +0.727 & +0.688 & +0.751 \\
$A \times 10^2$              & +0.126        & +0.061 & +0.062 & +0.259 & -0.149 & -0.361 & +0.555 & +0.136 &+0.137 & +0.450 & +0.344 \\
$\alpha$                     &  0.894   &  0.760  & 0.851 &  0.813 & 0.891 &  0.631 &  1.271 &  0.831 &  0.940 &  0.939 &  0.807 \\
$[(2-F_{\rm bkg}) \Delta - \alpha f_S \delta] \times 10^2$
                             & $-0.170$ & $-0.193$ & $-0.178$ & $-0.157$ & $-0.270$ & $-0.370$ & $-0.133$ & $-0.206$ & $-0.152$
                             & $-0.114$ & $-0.049$ \\
$f_S$                        &  0.536   &  0.583  & 0.557 &  0.516 & 0.509 &  0.560 &  0.472 &  0.534 &  0.493 &  0.537 &  0.536 \\
$F_{\rm bkg}$                &  0.389   &  0.336  & 0.365 &  0.384 & 0.405 &  0.338 &  0.627 &  0.374 &  0.407 &  0.436 &  0.325 \\
\hline
$\aslb \times 10^2$
                             & -0.787 & $-0.803$ & $-0.891$  & $-0.600$ & $-0.906$ & $-0.708$ & $-1.138$ & $-0.584$ & $-0.986$
                             & $-0.379$ & $-0.654$ \\
$\sigma(\aslb) \times 10^2$ (stat)
                             &  0.172   & 0.278    & 0.204 &  0.335 & 0.207 &  0.220 &  0.365 &  0.224 &  0.302 &  0.263 &  0.254 \\
$\sigma(\aslb) \times 10^2$ (syst)
                             &  0.093   & 0.125    & 0.128 &  0.188 & 0.107 & 0.104  &  0.323 &  0.108 &  0.135 &  0.209 &  0.103 \\
Significance                 &          & 0.007    & 0.742 & 0.567 &  1.029 &  0.525 &  1.022 &  1.236 &  0.960 &  1.537 &  1.120
\end{tabular}
\end{ruledtabular}
\end{table*}

\begin{table*}
\caption{\label{tab21}
Measured asymmetry $\aslb$ for the reference selection (column \textsf{Ref}) and
for samples used in Tests \textsf{I -- P}.
}
\begin{ruledtabular}
\newcolumntype{A}{D{A}{\pm}{-1}}
\newcolumntype{B}{D{B}{.}{-1}}
\begin{tabular}{lrrrrrrrrr}
& \textsf{Ref} & \textsf{I} & \textsf{J} & \textsf{K} & \textsf{L} & \textsf{M}
& \textsf{N} & \textsf{O} & \textsf{P}
\\ \hline
$N(\mu \mu) \times 10^{-6}$  &  6.019 &  4.428 &  3.504 &  2.928 &  2.741 &  4.259 &  3.709 &  2.724 &  2.440  \\
$a \times 10^2$              & +0.688 & +0.672 & +0.691 & +0.711 & +0.761 & +0.501 & +0.802 & +0.688 & +0.688  \\
$ A \times 10^2$             & +0.126 & +0.250 & +0.160 & +0.118 & +0.216 & --0.033 & +0.262 & +0.245 & +0.272 \\
$\alpha$                     &  0.894 &  0.908 &  0.817 &  0.872 &  0.825 &  0.702 &  0.908 &  0.941 &  0.898  \\
$[(2-F_{\rm bkg}) \Delta - \alpha f_S \delta] \times 10^2$
                             & $-0.170$ & $-0.209$ & $-0.187$ & $-0.221$ & $-0.214$ & $-0.187$ & $-0.150$ & $-0.126$
                             & $-0.122$ \\
$f_S$                        &  0.536 &  0.514 &  0.555 &  0.556 &  0.570 &  0.519 &  0.514 &  0.536 &  0.536  \\
$F_{\rm bkg}$                &  0.389 &  0.414 &  0.352 &  0.363 &  0.333 &  0.402 &  0.428 &  0.408 &  0.395  \\
\hline
$\aslb \times 10^2$
                             & $-0.787$ & $-0.925$ & $-0.569$ & $-0.847$ & $-0.430$ & $-0.761$ & $-0.774$ & $-0.809$
                             & $-0.689$ \\
$\sigma(\aslb) \times 10^2$ (stat) 
                             &  0.172 &  0.204 &  0.202 &  0.224 &  0.260 &  0.207 &  0.221 &  0.247 &  0.253 \\
$\sigma(\aslb) \times 10^2$ (syst) 
                             &  0.093 &  0.115 &  0.100 &  0.122 &  0.117 &  0.110 &  0.118 &  0.129 &  0.128 \\
Significance                 &        &  1.245 &  1.672 &  0.377 &  1.678 &  0.441 &  0.186 &  0.120 &  0.497
\end{tabular}
\end{ruledtabular}
\end{table*}

\begin{figure}
\begin{center}
\includegraphics[width=0.50\textwidth]{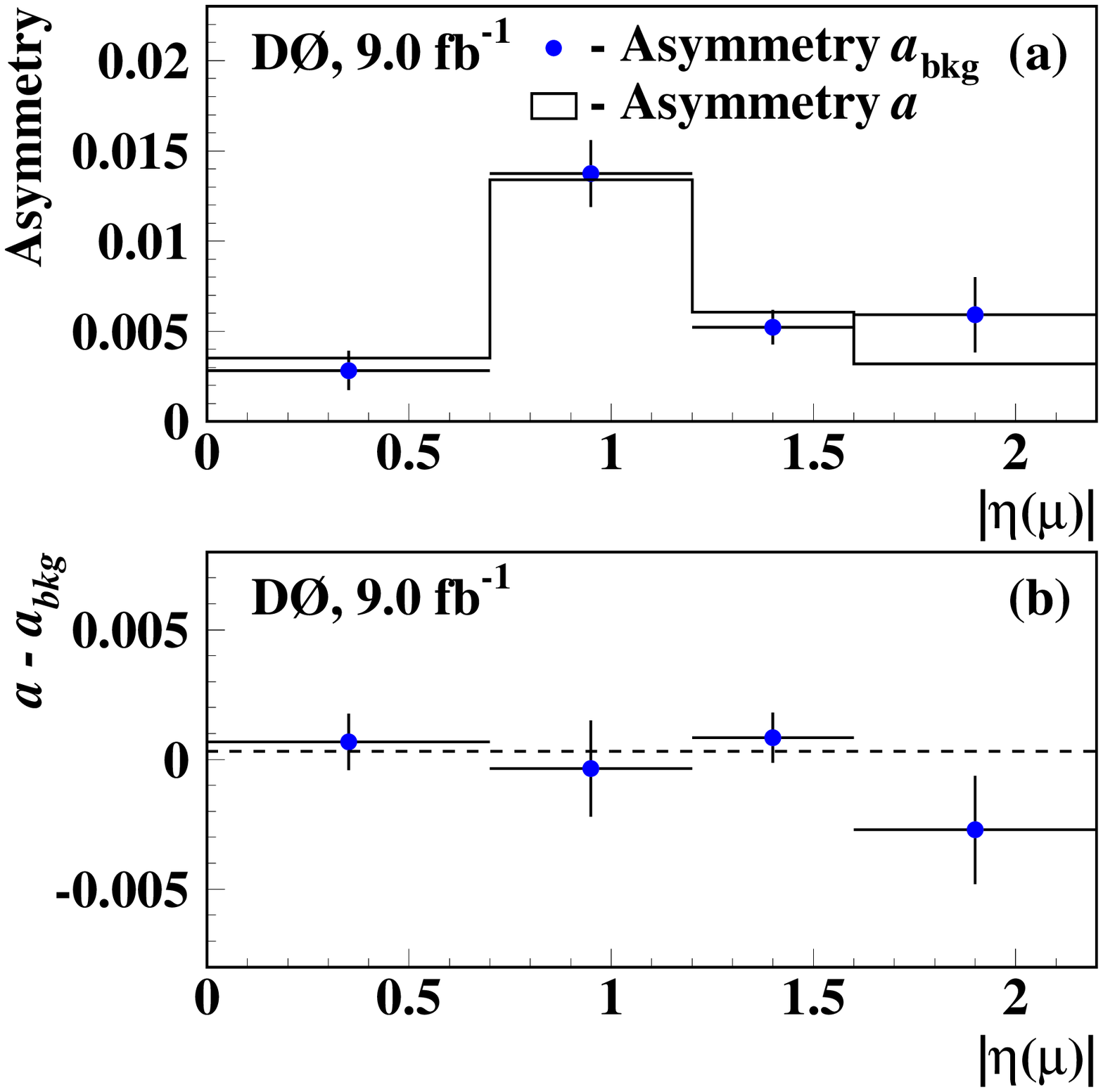}
\caption{(color online). (a)
The asymmetry $a_{\rm bkg}$ (points with error
bars representing the total uncertainties), as expected from the measurements of the fractions and
asymmetries of the background processes, is compared to the measured asymmetry $a$
of the inclusive muon sample (shown as a histogram, since the statistical uncertainties
are negligible) as a function of the absolute value of muon pseudorapidity $|\eta(\mu)|$.
The asymmetry from {\sl CP} violation is negligible
compared to the background in the inclusive muon sample.
(b)~The difference $a - a_{\rm bkg}$. The horizontal dashed line shows the mean value.
}
\label{fig10-eta}
\end{center}
\end{figure}

We also compare the dependence on the muon pseudorapidity $\eta(\mu)$
of the observed and expected charge asymmetry in the inclusive
muon sample. We repeat the analysis
procedure, but measure all background contributions as a function of $|\eta(\mu)|$.
The result of this comparison is shown in Fig.~{\ref{fig10-eta}}.
The dependence on $|\eta(\mu)|$ is correctly
described by the background asymmetry. There is good agreement between these
two quantities, with a $\chi^2/{\rm d.o.f.}= 2.8/4$.
This is consistent with our expectation that
the contribution of $\aslb$ in the inclusive muon charge asymmetry is
overwhelmed by background.

\begin{figure}
\begin{center}
\includegraphics[width=0.50\textwidth]{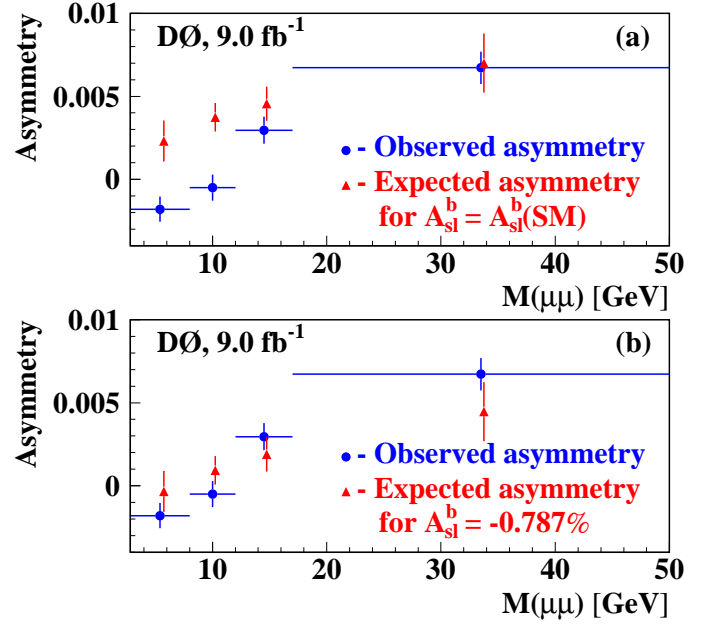}
\caption{(color online). The observed and expected like-sign dimuon charge asymmetries
in bins of dimuon invariant mass. The expected asymmetry
is shown for (a) $\aslb = \aslb({\rm SM})$ and (b) $\aslb = -0.787\%$.
}
\label{a-mmm}
\end{center}
\end{figure}

Figure~\ref{a-mmm} shows the observed and expected uncorrected like-sign dimuon charge
asymmetry as a function of the dimuon invariant mass. The expected asymmetry is computed
using Eq.~(\ref{dimuon_A}) and the measured parameters of sample composition and asymmetries.
As in Ref.~\cite{PRD}, we compare the expected uncorrected asymmetry using two
assumptions for $\aslb$. In Fig.~\ref{a-mmm}(a) the observed asymmetry is compared
to the expectation for the SM value of $\aslb({\rm SM}) = -0.028\%$,
while Fig.~\ref{a-mmm}(b) shows the expected asymmetry
for $\aslb = -0.787\%$.
Large discrepancies between the observed and expected asymmetries can be observed
for $\aslb = \aslb({\rm SM})$, while good agreement is obtained for the measured
$\aslb$ value corresponding to Eq.~(\ref{ah3}).
The observed asymmetry changes as a function of dimuon invariant mass,
and the expected asymmetry tracks this effect when $\aslb = -0.787\%$.
This dependence of the asymmetry on invariant
mass of the muon pair is a function of the production mechanism of
the particles involved and of their decays. The agreement between
the observed and expected asymmetries indicates that the physics
leading to the observed asymmetry can be described by contributions from the
background and from decays of $b$ hadrons.

We also measure the mean mixing probability using the ratio of like-sign and opposite-sign dimuon
events. The background contribution in both samples is obtained using the method presented
in this Article. The measured mean mixing probability is found to be consistent
with the world average value \cite{hfag}.

We conclude that our method of analysis provides a consistent description
of the dimuon charge asymmetry for a wide range of input parameters, as well as
for significantly modified selection criteria.


\section{Comparison with simulation}
\label{sec_sim}

The measurement of the background fractions is based on data, and the input from simulation
is limited to the
ratio of multiplicities $n_\pi / n_K$ and $n_p / n_K$ in $p \bar p$ interactions
\cite{PRD}. Nevertheless, it is instructive to compare the results obtained in
data and in simulation.
Such a comparison is shown in Table \ref{tab15}. The simulation used in this analysis
is described in Ref.~\cite{PRD}. All quantities measured in simulation
are obtained using the information on the generated processes.
All uncertainties in the second and third
columns are statistical. The difference between the values obtained in
data and simulation is given in the fourth column and includes the systematic uncertainties.
The agreement between the measured and simulated quantities
is satisfactory.
The excellent agreement between the mean values of $R_K$, which is one of the most essential
quantities of this measurement and for which many systematic uncertainties cancel, is
especially notable:
\begin{eqnarray}
R_K ({\rm data}) & = & 0.856 \pm 0.020~({\rm stat}) \pm 0.026~({\rm syst}), \nonumber \\
R_K ({\rm MC}) & = & 0.901 \pm 0.086~({\rm MC~stat}).
\end{eqnarray}
This comparison provides support for the validity of the presented measurement.

\begin{table}
\caption{\label{tab15}
Comparison of background fractions measured in data and in simulation.
Only the statistical uncertainties are given in the second and third column.
The difference between data and simulation is given in the fourth column
and includes both statistical and systematic uncertainties.
}
\begin{ruledtabular}
\newcolumntype{A}{D{A}{\pm}{-1}}
\newcolumntype{B}{D{B}{-}{-1}}
\begin{tabular}{lAAA}
Quantity &  \multicolumn{1}{c}{Data} &
\multicolumn{1}{c}{Simulation} & \multicolumn{1}{c}{Difference}\\
\hline
$f_K \times 10^2$ & 15.96 \ A \ 0.24 & 14.31 \ A \ 0.06 & +1.65 \ A \ 2.55 \\
$f_\pi \times 10^2$ & 30.01 \ A \ 1.60 & 29.82 \ A \ 0.09 & +0.19 \ A \ 5.15 \\
$f_p \times 10^2$ & 0.38 \ A \ 0.17 & 1.07 \ A \ 0.02 & -0.69 \ A \ 0.60 \\
$F_K \times 10^2$ & 13.78 \ A \ 0.42 & 12.89 \ A \ 1.32 & +0.89 \ A \ 2.26 \\
$F_\pi \times 10^2$ & 24.81 \ A \ 1.38 & 25.88 \ A \ 1.86 & -1.07 \ A \ 4.36 \\
$F_p \times 10^2$ & 0.35 \ A \ 0.14 & 1.29 \ A \ 0.39 & -0.94 \ A \ 0.72 \\ \hline
$f_S \times 10^2$ & 53.65 \ A \ 1.74 & 54.79 \ A \ 0.14 & -1.14 \ A \ 7.11 \\
$F_{\rm bkg} \times 10^2$ & 38.94 \ A \ 1.89 & 40.01 \ A \ 2.31 & -1.07 \ A \ 6.21 \\
$R_K \times 10^2$         & 85.62 \ A \ 1.98 & 90.08 \ A \ 8.60 & -4.46 \ A \ 9.74
\end{tabular}
\end{ruledtabular}
\end{table}

\section{Dependence of Asymmetry $\bm{\aslb}$ on muon impact parameter}
\label{sec_asip}

\begin{figure}
\begin{center}
\includegraphics[width=0.50\textwidth]{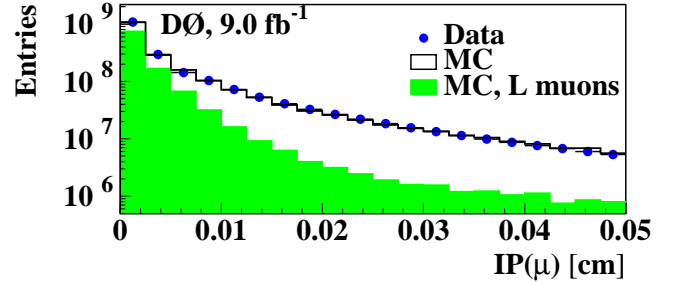}
\caption{(color online). The muon IP distribution in the inclusive muon sample
(bullets).
The solid line represents the muon IP distribution in simulation.
The shaded histogram is the contribution from $L$ muons in simulation.}
\label{fig-ip}
\end{center}
\end{figure}

The asymmetry $\aslb$ is produced by muons from direct semi-leptonic decays of $b$ quarks.
A distinctive feature of these muons is the large impact parameter of their trajectories
with respect to the primary vertex \cite{Gronau-Rosner,impact}.
The simulation shows that the
dominant source of background from $L$ muons corresponds to charged hadrons
produced in the primary interactions that then decay to muons, and the tracks of
such muons have small impact parameters if the decay is outside the tracking volume.
Figure~\ref{fig-ip} shows the muon IP distribution in data and in simulation.
The shaded histogram shows the contribution from $L$ muons in simulation,
which decreases significantly for increasing values of the muon IP. The background
can therefore be significantly suppressed by selecting muons with large impact parameter.

To verify the origin of the observed charge asymmetry, we perform several
complementary measurements.
We require the muon IP to be larger or smaller than 120 $\mu$m.
For events in the like-sign dimuon sample, we require that both muons satisfy
these conditions. These measurements are denoted
as $IP_{>120}$ and $IP_{<120}$, respectively. The selected threshold of 120 $\mu$m
can be compared with the spread in the crossing point of the colliding beams in the
Tevatron collider, and with the precision of $p \bar p$ vertex reconstruction,
which are about 30 $\mu$m and 15 $\mu$m, respectively,
in the plane perpendicular to the beam axis. The chosen value of 120~$\mu$m gives
the minimal uncertainty on $\asld$ and $\asls$ defined in Eq.~(\ref{Ab_7}).

In total, $0.356 \times 10^9$ muons in the inclusive muon sample
and $0.714 \times 10^6$ events in the like-sign dimuon sample are selected for
the $IP_{>120}$ measurement. Events are subject to the same analysis
as for the entire sample, except that the ratio $R_K(\ks)$ is not used because of
insufficient $\ks \to \pi^+ \pi^-$ decays in the dimuon sample.
Background asymmetries should not depend on the muon IP, and
we verified that the difference in kaon asymmetry for the whole sample and
the $IP_{>120}$ events agree: $a_K(IP_{>120}) - a_K({\rm all}) = (-1.6 \pm 1.5)\%$.
We therefore use the values
given in Tables \ref{tab6} and \ref{tab8}. All other measured quantities are given
in Table \ref{tab16}. The background fractions are strongly suppressed
in the $IP_{>120}$ sample, and their influence on the measurement of $\aslb$
is significantly smaller.
Using these values, we obtain for the inclusive muon sample
\begin{equation}
\aslb (IP_{>120}) = (-0.422 \pm 0.240~({\rm stat}) \pm 0.121~({\rm syst}))\%,
\label{ah1_ip}
\end{equation}
and for the like-sign dimuon sample
\begin{equation}
\aslb (IP_{>120})= (-0.818 \pm 0.342~({\rm stat}) \pm 0.067~({\rm syst}))\%.
\label{ah2_ip}
\end{equation}

\begin{table}
\caption{\label{tab16}
Input quantities for the measurement of $\aslb$ using muons with IP above
and below 120~$\mu m$. Only statistical uncertainties are given.
}
\begin{ruledtabular}
\newcolumntype{A}{D{A}{\pm}{-1}}
\newcolumntype{B}{D{B}{-}{-1}}
\begin{tabular}{lAA}
Quantity &  \multicolumn{1}{c}{$IP >120~\mu$m} & \multicolumn{1}{c}{$IP < 120~\mu$m} \\
\hline
$f_K \times 10^2$ & 5.19\ A \ 0.37 & 17.64\ A \ 0.27 \\
$f_\pi \times 10^2$ & 5.65\ A \ 0.40 & 34.72\ A \ 1.86 \\
$f_p \times 10^2$ & 0.05\ A \ 0.03 &  0.45\ A \ 0.20 \\
$F_K \times 10^2$ & 4.48\ A \ 4.05 &  21.49\ A \ 0.62 \\
$F_\pi \times 10^2$ & 4.43\ A \ 3.95 & 40.47\ A \ 2.26 \\
$F_p \times 10^2$ & 0.03\ A \ 0.05 &  0.59\ A \ 0.23 \\ \hline
$f_S \times 10^2$ & 89.11\ A \ 0.88 & 47.18\ A \ 2.03 \\
$F_{\rm bkg} \times 10^2$ &  8.94\ A \ 8.26 & 62.56\ A \ 3.07 \\
$F_{SS} \times 10^2$ & 91.79\ A \ 7.65 & 53.66\ A \ 2.68 \\ \hline
$a \times 10^2$   & -0.014\ A \ 0.005 & +0.835\ A \ 0.002 \\
$a_{\rm bkg} \times 10^2$   & +0.027\ A \ 0.023 & +0.864\ A \ 0.049 \\
$A \times 10^2$   & -0.529 \ A \ 0.120 & +0.555    \ A \ 0.060 \\
$A_{\rm bkg} \times 10^2$   & -0.127 \ A \ 0.093 & +0.829 \ A \ 0.077 \\
\hline
$C_\pi$           & 0.70\ A \ 0.05 & 0.95\ A \ 0.02  \\
$C_K$             & 0.39\ A \ 0.06 & 0.98\ A  \ 0.01 \\
$F_{LL}/(F_{LL}+F_{SL})$ & 0.089\ A \ 0.062 & 0.350\ A \ 0.029 \\
$c_b$ & 0.109\ A \ 0.011 & 0.038\ A \ 0.007 \\
$C_b$ & 0.526\ A \ 0.037 & 0.413\ A \ 0.032
\end{tabular}
\end{ruledtabular}
\end{table}

We obtain the final value of $\aslb (IP_{>120}) $
using the linear combination of Eq.~(\ref{aprime}), and
select the value of $\alpha$ to minimize the total uncertainty on $\aslb$,
which corresponds to $\alpha = -9.29$.
The combination for a negative value of $\alpha$ is equivalent to the weighted average of
Eqs.~(\ref{ah1_ip}) and~(\ref{ah2_ip})
taking into account the correlation of uncertainties
(see Appendix \ref{app1} for more details). The corresponding asymmetry $\aslb$ is found to be
\begin{equation}
\aslb (IP_{>120})= (-0.579 \pm 0.210~({\rm stat}) \pm 0.094~({\rm syst}))\%.
\label{ah3_ip}
\end{equation}
The contributions to the uncertainties in Eqs.~(\ref{ah1_ip} -- \ref{ah3_ip}) are given
in Table \ref{tab17}.

\begin{table}
\caption{\label{tab17}
Sources of uncertainty on $\aslb (IP_{>120})$ in Eqs.~(\ref{ah1_ip}),
(\ref{ah2_ip}), and~(\ref{ah3_ip}). The first nine rows contain statistical uncertainties,
and the next four rows contain systematic uncertainties.
}
\begin{ruledtabular}
\newcolumntype{A}{D{A}{\pm}{-1}}
\newcolumntype{B}{D{B}{-}{-1}}
\begin{tabular}{cccc}
Source & $\delta(\aslb)\times 10^2$ & $\delta(\aslb)\times 10^2$ &
$\delta(\aslb) \times 10^2$ \\
       & Eq.~(\ref{ah1_ip}) & Eq.~(\ref{ah2_ip}) & Eq.~(\ref{ah3_ip}) \\
\hline
$A$ or $a$ (stat)       & 0.055 & 0.244 & 0.093 \\
$f_K$ (stat)            & 0.048 & 0.031 & 0.058 \\
$R_K$ (stat)            &  N/A    & 0.244 & 0.074 \\
$P(\pitomu)/P(\ktomu)$  & 0.007 & 0.004 & 0.006 \\
$P(\ptomu)/P(\ktomu)$   & 0.012 & 0.004 & 0.010 \\
$A_K$                   & 0.023 & 0.012 & 0.017 \\
$A_\pi$                 & 0.037 & 0.009 & 0.026 \\
$A_p$                   & 0.025 & 0.007 & 0.019 \\
$\delta$ or $\Delta$    & 0.210 & 0.075 & 0.157 \\
\hline
$f_K$ (syst)            & 0.112 & 0.027 & 0.083 \\
$R_K$ (syst)            &   N/A   & 0.014 & 0.007 \\
$\pi$, $K$, $p$
multiplicity            & 0.016 & 0.016 & 0.016 \\
$c_b$ or $C_b$          & 0.043 & 0.057 & 0.041 \\
\hline
Total statistical       & 0.240 & 0.342 & 0.210 \\
Total systematic        & 0.121 & 0.067 & 0.094 \\
Total                   & 0.269 & 0.348 & 0.230
\end{tabular}
\end{ruledtabular}
\end{table}

\begin{figure}
\begin{center}
\includegraphics[width=0.50\textwidth]{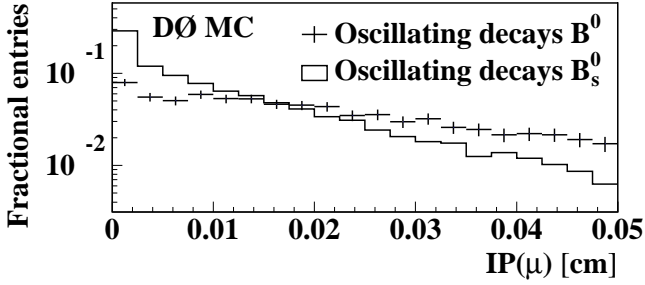}
\caption{(color online). The normalized IP distribution for muons produced in oscillating decays of
$\Bd$ mesons (points with error bars) and $\Bs$ mesons (solid histogram) in simulation. }
\label{fig-iposc}
\end{center}
\end{figure}

From the known frequencies of oscillations, $\Delta M_q/ 2 \pi$ $(q = d,s)$,
the period of oscillation for the $\Bd$ meson is many times longer than its lifetime
so that the mixing probability of $\Bd$ mesons effectively increases with long decay
lengths and large impact parameters.  The $\Bs$ meson oscillates a number of times
within its lifetime so that it is ``fully mixed" for any appreciable impact parameter
requirement. As a result, the fraction of $\Bd$ mesons that have oscillated
into the other flavor is increased in the sample with large muon impact parameter.
This behavior is demonstrated in Fig.~\ref{fig-iposc}, which
shows the normalized IP distributions for muons produced in oscillating decays of
$\Bd$ and $\Bs$ mesons in simulation. The contribution of the $\asld$ asymmetry in
$\aslb$ is therefore enhanced in the sample with a large muon IP. From simulation,
the mixing probability of $\Bd$ meson in the $IP_{>120}$ sample is determined to be
\begin{equation}
\chi_d (IP_{>120}, {\rm MC}) = 0.342 \pm 0.004,
\end{equation}
with the uncertainty limited by the number of simulated events. This value can be
compared to the input to the simulation for the $\Bd$ mixing probability
integrated over time, $\chi_d = 0.1864 \pm 0.0022$ \cite{hfag}.
The coefficients $C_d$ and $C_s$ in Eq.~(\ref{Ab_7})
for the $IP_{>120}$ selection become
\begin{eqnarray}
C_d(IP_{>120}) & = & 0.728 \pm 0.018, \nonumber \\
C_s(IP_{>120}) & = & 0.272 \pm 0.018.
\label{Ab_9}
\end{eqnarray}
The value of $\aslb (IP_{>120})$ should therefore be reduced relative to the value
for the full dimuon sample, if the contribution of $\asls$ dominates the asymmetry $\aslb$.

The measurement of $IP_{<120}$ is performed using
$1.687 \times 10^9$ muons in the inclusive muon sample
and $2.857 \times 10^6$ events in the like-sign dimuon sample. Exactly the same procedure
is applied as for the main measurement, using the background and muon reconstruction asymmetries
given in Tables \ref{tab6} and \ref{tab8}. All other quantities are given
in Table \ref{tab16}. The background fractions are significantly increased
in the samples with small muon IP, thereby increasing the uncertainties
related to the background description (Table \ref{tab23}).

\begin{table}
\caption{\label{tab23}
Sources of uncertainty on $\aslb (IP_{<120})$ in Eqs.~(\ref{ah4_ip}),
(\ref{ah5_ip}), and~(\ref{ah6_ip}). The first nine rows contain statistical uncertainties,
the next four rows contain systematic uncertainties.
}
\begin{ruledtabular}
\newcolumntype{A}{D{A}{\pm}{-1}}
\newcolumntype{B}{D{B}{-}{-1}}
\begin{tabular}{cccc}
Source & $\delta(\aslb)\times 10^2$ & $\delta(\aslb)\times 10^2$ &
$\delta(\aslb) \times 10^2$ \\
       & Eq.~(\ref{ah4_ip}) & Eq.~(\ref{ah5_ip}) & Eq.~(\ref{ah6_ip}) \\
\hline
$A$ or $a$ (stat)       & 0.136 & 0.233 & 0.285 \\
$f_K$ (stat)            & 1.059 & 0.173 & 0.082 \\
$R_K$ (stat)            &  N/A    & 0.141 & 0.155 \\
$P(\pitomu)/P(\ktomu)$  & 0.388 & 0.060 & 0.026 \\
$P(\ptomu)/P(\ktomu)$   & 0.699 & 0.064 & 0.004 \\
$A_K$                   & 0.986 & 0.123 & 0.089 \\
$A_\pi$                 & 1.727 & 0.165 & 0.075 \\
$A_p$                   & 1.261 & 0.123 & 0.050 \\
$\delta$ or $\Delta$    & 0.606 & 0.107 & 0.071 \\
\hline
$f_K$ (syst)            & 4.951 & 0.508 & 0.034 \\
$R_K$ (syst)            &   N/A   & 0.286 & 0.307 \\
$\pi$, $K$, $p$
multiplicity            & 0.137 & 0.034 & 0.025 \\
$c_b$ or $C_b$          & 0.305 & 0.087 & 0.093 \\
\hline
Total statistical       & 2.774 & 0.439 & 0.366 \\
Total systematic        & 4.962 & 0.590 & 0.323 \\
Total                   & 5.685 & 0.735 & 0.488
\end{tabular}
\end{ruledtabular}
\end{table}

Using these values we obtain from the inclusive muon sample
\begin{equation}
\aslb (IP_{<120}) = (-1.65 \pm 2.77~({\rm stat}) \pm 4.96~({\rm syst}))\%,
\label{ah4_ip}
\end{equation}
and from the like-sign dimuon sample
\begin{equation}
\aslb (IP_{<120})=  (-1.17 \pm 0.44~({\rm stat}) \pm 0.59~({\rm syst}))\%.
\label{ah5_ip}
\end{equation}
The measurement using the linear combination given in Eq.~(\ref{aprime})
is performed with $\alpha = +1.27$, which minimizes the total
uncertainty on $\aslb$. The value of $\aslb$ is found to be
\begin{equation}
\aslb (IP_{<120})= (-1.14 \pm 0.37~({\rm stat}) \pm 0.32~({\rm syst}))\%.
\label{ah6_ip}
\end{equation}
The mean mixing probability $\chi_d$ in the $IP_{<120}$ sample obtained in simulation
is found to be
\begin{equation}
\chi_d (IP_{<120}, MC) = 0.084 \pm 0.002,
\end{equation}
and the coefficients $C_d$ and $C_s$ in Eq.~(\ref{Ab_7})
for the $IP_{<120}$ selection are
\begin{eqnarray}
C_d(IP_{<120}) & = & 0.397 \pm 0.022, \nonumber \\
C_s(IP_{<120}) & = & 0.603 \pm 0.022.
\label{Ab_10}
\end{eqnarray}

\begin{figure}
\begin{center}
\includegraphics[width=0.50\textwidth]{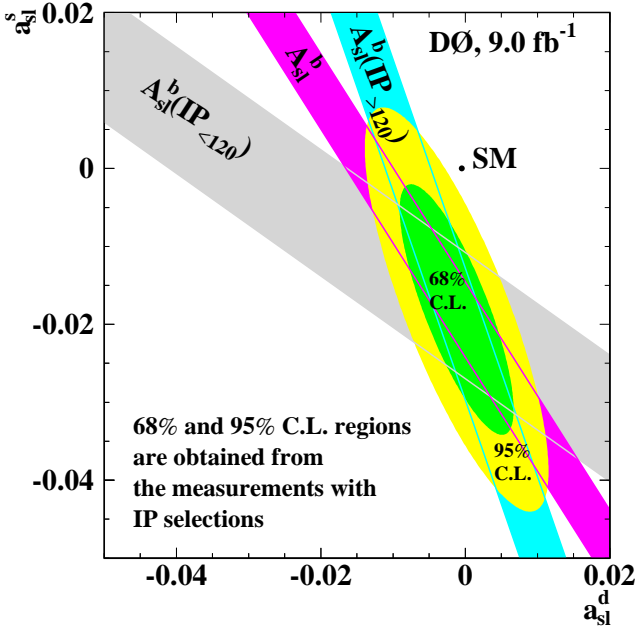}
\caption{(color online). Measurements of $\aslb$ with different muon IP
selections in the ($\asld,\asls$) plane.
The bands represent the $\pm 1$ standard deviation uncertainties on
each individual measurement. The ellipses represent the 68\% and 95\%
two-dimensional C.L. regions, respectively,
of $\asls$ and $\asls$ values obtained from the measurements with IP selections. }
\label{fig-ads2}
\end{center}
\end{figure}

The measurements with $IP_{<120}$ and $IP_{>120}$ use independent data samples, and
the dependence of $\aslb$ on $\asld$ and $\asls$ is different for the $IP_{<120}$
and $IP_{>120}$ samples. The measurements given in Eqs.~(\ref{ah3_ip}) and
(\ref{ah6_ip}) can therefore be combined to obtain the values of $\asld$ and $\asls$,
taking into account the correlation among different
sources of uncertainty. All uncertainties in
Tables \ref{tab17}, \ref{tab23},
except the statistical uncertainties on $a$, $A$, $f_K$, $R_K$,
$P(\pi \to \mu)/P(K \to \mu)$, and $P(p \to \mu)/P(K \to \mu)$ are treated
as fully correlated. The values of $\asld$ and $\asls$ extracted are
\begin{eqnarray}
\asld & = & (-0.12 \pm 0.52) \%, \nonumber \\
\asls & = & (-1.81 \pm 1.06) \%.
\end{eqnarray}
The correlation $\rho_{ds}$ between these two quantities is
\begin{equation}
\rho_{ds} = -0.799.
\end{equation}
The uncertainty on $\asld$ and $\asls$ obtained in this study is comparable
with that obtained from the direct measurements.
Figure~\ref{fig-ads2} presents the results of the IP study in the
($\asld,\asls$) plane together with
the result (\ref{ah3}) of the $\aslb$ measurement
using all like-sign dimuon events.
The ellipses represent the 68\% and 95\%
two-dimensional C.L. regions, respectively,
of $\asls$ and $\asls$ values obtained from the measurements with IP selections.

We also performed four additional measurements with IP thresholds
of 50 $\mu$m and 80 $\mu$m. They are denoted as $IP_{<50}$,
$IP_{>50}$, $IP_{<80}$, and $IP_{>80}$, respectively.
The input quantities for these measurements are presented in Tables \ref{tab19}
and \ref{tab24}. The $\aslb$ values in the inclusive and like-sign dimuon
samples and their combinations are given in Table \ref{tab25}. The mean mixing
probability $\chi_d$ for all these measurement is obtained through simulation.
The results are presented in Table \ref{tab26}, together with the corresponding
coefficients $C_d$ and $C_s$.

As for the combinations of the $IP_{<120}$ and $IP_{>120}$ samples,
the measurements with $IP_{<50}$ and $IP_{>50}$ samples, as well as
with $IP_{<80}$ and $IP_{>80}$ samples, can be combined to determine
the values of $\asld$ and $\asls$ (Table\ \ref{tab27}).
The measurements with different IP thresholds are consistent
with each other within two standard deviations taking into account the correlation
between the uncertainties.

%


We conclude that the observed dependence of the like-sign dimuon charge
asymmetry on muon IP is consistent with the hypothesis that it
has its origin from semi-leptonic $b$-hadron decays.
The contributions of $\asld$ and $\asls$ to $\aslb$ can be
 determined separately by dividing the sample according to the muon IP,
 although the uncertainties on the values of $\asld$ and $\asls$ do not
 allow for the definitive conclusion that the deviation of $\aslb$ from
 its SM prediction is dominated from the $\asls$ asymmetry.


\begin{table}
\caption{\label{tab19}
Input quantities for the measurement of $\aslb$ using muons with IP above
50 $\mu m$, 80 $\mu$m and 120 $\mu$m, respectively. Only statistical uncertainties are given.
}
\begin{ruledtabular}
\newcolumntype{A}{D{A}{\pm}{-1}}
\newcolumntype{B}{D{B}{-}{-1}}
\begin{tabular}{lAAA}
Quantity &  \multicolumn{1}{c}{$IP >50~\mu$m} & \multicolumn{1}{c}{$IP > 80~\mu$m} &
            \multicolumn{1}{c}{$IP > 120~\mu$m}\\
\hline
$f_K \times 10^2$ & 6.47\ A \ 0.18 & 5.38\ A \ 0.24 & 5.19\ A \ 0.37 \\
$f_\pi \times 10^2$ &10.42\ A \ 0.47 & 7.24\ A \ 0.38 & 5.65\ A \ 0.40\\
$f_p \times 10^2$ & 0.11\ A \ 0.05 &  0.07\ A \ 0.03 & 0.05\ A \ 0.03  \\
$F_K \times 10^2$ & 6.31\ A \ 1.73 &  4.79\ A \ 2.59 & 4.48\ A \ 4.05 \\
$F_\pi \times 10^2$ & 9.51\ A \ 2.36 & 6.39\ A \ 2.95 & 4.43\ A \ 3.95 \\
$F_p \times 10^2$ & 0.11\ A \ 0.06 &  0.03\ A \ 0.04 & 0.03\ A \ 0.05 \\ \hline
$f_S \times 10^2$ & 82.99\ A \ 0.81 & 87.32\ A \ 0.74 & 89.11\ A \ 0.88 \\
$F_{\rm bkg} \times 10^2$ & 15.91\ A \ 4.38 &11.39\ A \ 6.10 & 8.94\ A \ 8.26 \\
$F_{SS} \times 10^2$ & 85.63\ A \ 3.74 & 89.88\ A \ 5.10 & 91.79\ A \ 7.65 \\
\hline
$a \times 10^2$   & +0.134\ A \ 0.004 & +0.035\ A \ 0.005 & -0.014\ A \ 0.005 \\
$a_{\rm bkg} \times 10^2$   & +0.146\ A \ 0.024 & +0.068\ A \ 0.023 & +0.027\ A \ 0.023 \\
$A \times 10^2$   & -0.302 \ A \ 0.079 & -0.386 \ A \ 0.094 & -0.529 \ A \ 0.120 \\
$A_{\rm bkg} \times 10^2$   & -0.043 \ A \ 0.071 & -0.139 \ A \ 0.083 & -0.127 \ A \ 0.093 \\
\hline
$C_\pi$           & 0.81\ A \ 0.03 & 0.75\ A \ 0.05  & 0.70  \ A \ 0.05 \\
$C_K$             & 0.66\ A \ 0.03 & 0.52\ A  \ 0.05 & 0.39  \ A \ 0.06 \\
$\frac{F_{LL}}{(F_{LL}+F_{SL})}$ & 0.108\ A \ 0.038 & 0.125\ A \ 0.060 & 0.089\ A \ 0.062 \\
$c_b$ & 0.084\ A \ 0.008 & 0.095\ A \ 0.009 & 0.109\ A \ 0.011 \\
$C_b$ & 0.496\ A \ 0.034 & 0.510\ A \ 0.034 & 0.526\ A \ 0.037
\end{tabular}
\end{ruledtabular}
\end{table}

\begin{table}
\caption{\label{tab24}
Input quantities for the measurement of $\aslb$ using muons with IP below
50 $\mu m$, 80 $\mu$m and 120 $\mu$m, respectively. Only statistical uncertainties are given.}
\begin{ruledtabular}
\newcolumntype{A}{D{A}{\pm}{-1}}
\newcolumntype{B}{D{B}{-}{-1}}
\begin{tabular}{lAAA}
Quantity &  \multicolumn{1}{c}{$IP < 50 \mu$m} & \multicolumn{1}{c}{$IP <80~\mu$m} &
       \multicolumn{1}{c}{$IP < 120~\mu$m}\\
\hline
$f_K \times 10^2$ &19.35\ A \ 0.33 &18.32\ A \ 0.30 &17.64\ A \ 0.27 \\
$f_\pi \times 10^2$ &37.58\ A \ 2.08 &34.34\ A \ 1.95 &34.72\ A \ 1.86\\
$f_p \times 10^2$ & 0.51\ A \ 0.22 &  0.48\ A \ 0.21 & 0.45\ A \ 0.20  \\
$F_K \times 10^2$ &28.03\ A \ 0.95 & 23.79\ A \ 0.74 &21.49\ A \ 0.62 \\
$F_\pi \times 10^2$ &51.72\ A \ 3.18 &44.26\ A \ 2.63 &40.47\ A \ 2.26 \\
$F_p \times 10^2$ & 0.77\ A \ 0.29 &  0.66\ A \ 0.25 & 0.59\ A \ 0.23 \\ \hline
$f_S \times 10^2$ & 42.56\ A \ 2.73 & 45.40\ A \ 2.13 & 47.18\ A \ 2.03 \\
$F_{\rm bkg} \times 10^2$ & 81.53\ A \ 4.30 &70.13\ A \ 3.52 &62.56\ A \ 3.07 \\
$F_{SS} \times 10^2$ & 43.42\ A \ 3.75 & 48.76\ A \ 2.84 & 53.66\ A \ 2.68 \\
\hline
$a \times 10^2$   & +0.953\ A \ 0.003 & +0.896\ A \ 0.003 & +0.835\ A \ 0.002 \\
$a_{\rm bkg} \times 10^2$   & +0.997\ A \ 0.056 & +0.916\ A \ 0.052 & +0.864\ A \ 0.049 \\
$A \times 10^2$   & +0.715    \ A \ 0.083 & +0.683 \ A \ 0.069 & +0.555 \ A \ 0.060 \\
$A_{\rm bkg} \times 10^2$   & +1.243 \ A \ 0.096 & +0.994 \ A \ 0.082 & +0.829  \ A \ 0.077 \\
\hline
$C_\pi$           & 0.97\ A \ 0.01 & 0.95\ A \ 0.02  & 0.95  \ A \ 0.02 \\
$C_K$             & 0.99\ A \ 0.01 & 0.98\ A  \ 0.01 & 0.98  \ A \ 0.01 \\
$\frac{F_{LL}}{(F_{LL}+F_{SL})}$ & 0.441\ A \ 0.050 & 0.369\ A \ 0.032 & 0.350\ A \ 0.029 \\
$c_b$ & 0.033\ A \ 0.007 & 0.035\ A \ 0.007 & 0.038\ A \ 0.007 \\
$C_b$ & 0.406\ A \ 0.032 & 0.406\ A \ 0.032 & 0.413\ A \ 0.032
\end{tabular}
\end{ruledtabular}
\end{table}

\begin{table}
\caption{\label{tab25}
Values of $\aslb$ with their statistical and systematic uncertainties obtained
for different IP selections.
}
\begin{ruledtabular}
\newcolumntype{A}{D{A}{\pm}{-1}}
\newcolumntype{B}{D{B}{-}{-1}}
\begin{tabular}{lcccc}
Selection & Sample & Central & \multicolumn{2}{c}{Uncertainty $\times 10^2$}  \\
          &        & value $\times 10^2$ & statistical & systematic \\
\hline
\multirow{3}{*}{All events}& $1\mu$ & --1.042  & 1.304 & 2.313  \\
                           & $2\mu$ & --0.808  & 0.202 & 0.222 \\
                           & comb.  & --0.787  & 0.172 & 0.093 \\ \hline
\multirow{3}{*}{$IP <50~\mu$m}& $1\mu$ & --3.244  & 4.101 & 7.466  \\
                           & $2\mu$ & --2.837     & 0.776 & 1.221 \\
                           & comb.  & --2.779     & 0.674 & 0.694 \\ \hline
\multirow{3}{*}{$IP > 50~\mu$m}& $1\mu$ & --0.171  & 0.343 & 0.311  \\
                           & $2\mu$ & --0.593      & 0.257 & 0.074 \\
                           & comb.  & --0.533      & 0.239 & 0.100 \\ \hline
\multirow{3}{*}{$IP < 80~\mu$m}& $1\mu$ & --1.293  & 3.282 & 5.841  \\
                           & $2\mu$ & -1.481   & 0.541 & 0.810 \\
                           & comb.  & -1.521   & 0.458 & 0.501 \\ \hline
\multirow{3}{*}{$IP > 80~\mu$m}& $1\mu$ & --0.388  & 0.280 & 0.179  \\
                           & $2\mu$ & --0.529      & 0.285 & 0.048 \\
                           & comb.  & --0.472      & 0.226 & 0.091 \\ \hline
\multirow{3}{*}{$IP < 120~\mu$m}& $1\mu$ & --1.654 & 2.774 & 4.962  \\
                           & $2\mu$ & --1.175      & 0.439 & 0.590 \\
                           & comb.  & --1.138      & 0.366 & 0.323 \\ \hline
\multirow{3}{*}{$IP > 120~\mu$m}& $1\mu$ & --0.422 & 0.240 & 0.121  \\
                           & $2\mu$ & --0.818      & 0.342 & 0.067 \\
                           & comb.  & --0.579      & 0.210 & 0.094 \\
\end{tabular}
\end{ruledtabular}
\end{table}


\begin{table}
\caption{\label{tab26}
Mean mixing probability ($\chi_d$) obtained
in simulation, and the coefficients $C_d$ and $C_s$ in Eq.~(\ref{Ab_7}), used for different
selections.
}
\begin{ruledtabular}
\newcolumntype{A}{D{A}{\pm}{-1}}
\newcolumntype{B}{D{B}{-}{-1}}
\begin{tabular}{lAAA}
Sample & \multicolumn{1}{c}{$\chi_d (MC)$} & \multicolumn{1}{c}{$C_d$} & \multicolumn{1}{c}{$C_s$} \\
\hline
$IP_{<50}$   & 0.059 \ A \ 0.002 & 0.316 \ A \ 0.021 & 0.684 \ A \ 0.021 \\
$IP_{<80}$   & 0.069 \ A \ 0.002 & 0.351 \ A \ 0.022 & 0.649 \ A \ 0.022 \\
$IP_{<120}$  & 0.084 \ A \ 0.002 & 0.397 \ A \ 0.022 & 0.603 \ A \ 0.022 \\
$IP_{>50}$   & 0.264 \ A \ 0.004 & 0.674 \ A \ 0.020 & 0.326 \ A \ 0.020 \\
$IP_{>80}$   & 0.299 \ A \ 0.004 & 0.701 \ A \ 0.019 & 0.299 \ A \ 0.019 \\
$IP_{>120}$  & 0.342 \ A \ 0.004 & 0.728 \ A \ 0.018 & 0.272 \ A \ 0.018
\end{tabular}
\end{ruledtabular}
\end{table}

\begin{table}
\caption{\label{tab27}
Measured values of $\asld$ and $\asls$ for different muon IP thresholds.
In each column, the measurements using the samples
with muon IP larger and smaller than the given
threshold are combined.
We also give the correlation $\rho_{ds}$ between $\asld$ and $\asls$.
}
\begin{ruledtabular}
\newcolumntype{A}{D{A}{\pm}{-1}}
\newcolumntype{B}{D{B}{-}{-1}}
\begin{tabular}{lAAA}
Quantity & \multicolumn{3}{c}{muon IP threshold} \\
         & \multicolumn{1}{c}{50 $\mu$m} & \multicolumn{1}{c}{80 $\mu$m} &
           \multicolumn{1}{c}{120 $\mu$m}\\
\hline
$\asld \times 10^2$ &  +1.51\ A \ 0.93 & +0.42\ A \ 0.68 & -0.12\ A \ 0.52 \\
$\asls \times 10^2$ &  -4.76\ A \ 1.79 & -2.57\ A \ 1.34 & -1.81\ A \ 1.06 \\
$\rho_{ds}$  & \multicolumn{1}{c}{$-0.912$} & \multicolumn{1}{c}{$-0.857$} & \multicolumn{1}{c}{$-0.799$} \\
\end{tabular}
\end{ruledtabular}
\end{table}


\section{Conclusions}
\label{conclusions}

We have presented an update to the previous
measurement \cite{PRD} of the anomalous like-sign dimuon charge asymmetry $\aslb$
with 9.0~fb$^{-1}$ of integrated luminosity.
The analysis has improved criteria for muon selection,
which provide a stronger background suppression and increase the
size of the like-sign dimuon sample.
A more accurate measurement
of the fraction of kaons that produce muons in the inclusive muon sample
($f_K$), and an additional measurement of the ratio
of such yields in like-sign dimuon to inclusive muon data ($R_K = F_K/f_K$) using
$\ks \to \pi^+ \pi^-$ decay have been performed.
This provides better precision of $R_K$, and an independent estimate of the systematic
uncertainty on this quantity. The value of the like-sign dimuon charge
asymmetry $\aslb$ in semi-leptonic $b$-hadron decays is found to be
\begin{equation}
\aslb = (-0.787 \pm 0.172~({\rm stat}) \pm 0.093~({\rm syst})) \%.
\end{equation}
This measurement disagrees with the prediction of the standard model by 3.9 standard
deviations and provides evidence for anomalously large $CP$ violation in semi-leptonic 
neutral $B$ decay. The residual charge asymmetry of like-sign dimuon events
after taking into account all background sources is found to be
\begin{equation}
A_{\rm res}  =  (-0.246 \pm 0.052~({\rm stat}) \pm 0.021~({\rm syst})) \%.
\end{equation}
It differs by 4.2 standard deviations from the standard model prediction.

Separation of the sample by muon impact parameter allows for separate extraction
of $\asld$ and $\asls$. We obtain
\begin{eqnarray}
\asld & = & (-0.12 \pm 0.52) \%, \nonumber \\
\asls & = & (-1.81 \pm 1.06) \%.
\end{eqnarray}
The correlation $\rho_{ds}$ between these two quantities is
\begin{equation}
\rho_{ds} = -0.799
.
\end{equation}
The uncertainties on $\asld$ and $\asls$ do not allow for the definitive
conclusion that $\asls$ dominates the value of $\aslb$.

Our results are consistent with the hypothesis that the anomalous like-sign
dimuon charge asymmetry arises from semi-leptonic $b$-hadron decays.
The significance of the difference of this measurement with
the SM prediction is not sufficient to claim observation
of physics beyond the standard model, but it has grown
compared to our previous measurement with a smaller data sample.


\begin{acknowledgments}
%
We thank the staffs at Fermilab and collaborating institutions,
and acknowledge support from the
DOE and NSF (USA);
CEA and CNRS/IN2P3 (France);
FASI, Rosatom and RFBR (Russia);
CNPq, FAPERJ, FAPESP and FUNDUNESP (Brazil);
DAE and DST (India);
Colciencias (Colombia);
CONACyT (Mexico);
KRF and KOSEF (Korea);
CONICET and UBACyT (Argentina);
FOM (The Netherlands);
STFC and the Royal Society (United Kingdom);
MSMT and GACR (Czech Republic);
CRC Program and NSERC (Canada);
BMBF and DFG (Germany);
SFI (Ireland);
The Swedish Research Council (Sweden);
and
CAS and CNSF (China).

\end{acknowledgments}

\appendix
\section{Combination of two measurements using $\bm{\alpha}$ scan}
\label{app1}

In this analysis, the value of $\aslb$ is obtained from the linear combination in Eq.~(\ref{aprime}). 
The parameter $\alpha$ is selected to minimize
the total uncertainty on $\aslb$, taking into account the correlation
among different contributions to the uncertainty on $\aslb$.
This procedure is equivalent to the standard procedure of taking a weighted average.

To demonstrate this, we consider a model in which
we obtain the quantity $x$ using two measurements $a$ and $A$.
Suppose that $a$ and $A$ depend linearly on $x$:
\begin{eqnarray}
a & = & k x + b, \nonumber \\
A & = & K x + B,
\end{eqnarray}
where $k$, $K$, $b$, and $B$ are parameters determined in the analysis, and
correspond to the measurement of $\aslb$.
Using the measurements of $a$ and $A$, we obtain two estimates of $x$:
\begin{eqnarray}
x_1 & = & (a-b)/k, \nonumber \\
x_2 & = & (A-B)/K.
\end{eqnarray}
We denote the uncertainties on $x_1$ and $x_2$ as $\sigma_1$
and $\sigma_2$, respectively.

Consider the case where the measurements of $a$ and $A$, as well as the
uncertainties $\sigma_1$ and $\sigma_2$, are statistically independent.
In this case, the value of $x$ can be obtained as a weighted average:
\begin{eqnarray}
x & = & (w_1 x_1 + w_2 x_2) / w, \nonumber \\
w_i & = & 1/\sigma_i^2,~ i=1,2, \nonumber \\
w   & = & 1/\sigma_1^2 + 1/\sigma_2^2.
\label{weigthed}
\end{eqnarray}

Consider another estimate of $x$ using the difference
\begin{equation}
A' = A - \alpha a,
\label{yprime}
\end{equation}
where $\alpha$ is a free parameter. The value of $x$ obtained
from Eq.~(\ref{yprime}) is
\begin{equation}
x = \frac{(A-B) - \alpha (a-b)}{K - \alpha k}.
\label{xalpha}
\end{equation}

Provided that the two measurements $a$ and $A$, as well as the uncertainties
$\sigma_1$ and $\sigma_2$ are statistically independent,
the minimal uncertainty on $x$ is obtained for
\begin{equation}
\alpha_{\rm min} = -(K \sigma_2^2) / (k \sigma_1^2).
\end{equation}
The central value and uncertainty on
$x$ obtained from Eq.~(\ref{xalpha}), with $\alpha = \alpha_{\rm min}$,
are exactly the same as the central value and uncertainty
obtained from the weighted average (\ref{weigthed}).
This case is similar to the combination (\ref{ah3_ip})
of the two measurements with $IP > 120~\mu$m that have reduced correlations.
The coefficient $\alpha$ in this case is negative, and its value depends
on the uncertainties $\sigma_1$ and $\sigma_2$.

Consider another extreme case, where $k=0$ and $B$ is fully correlated with $b$,
e.g., $B= C b$, where $C$ is a coefficient.
In this case, the value of $x$ obtained from Eq.~(\ref{xalpha}) is equal to
\begin{equation}
x = \frac{A - \alpha a - (C-\alpha) b}{K}.
\end{equation}
Provided that $\sigma(a) \ll \sigma(A)$, the minimal
uncertainty of $x$ is obtained for $\alpha_{\rm min} = C$.
This case corresponds to the measurement of Eq.~(\ref{ah3}) with the full data sample.
The value of $\alpha_{\rm min}$ is positive for $C > 0$.

These two examples demonstrate that the method of the $\alpha$ scan used in this
analysis is equivalent
to the weighted average of two measurements, taking into account
the correlation among different uncertainties.


\begin{thebibliography}{99}
\bibitem{charge}
Charge conjugate states are implied throughout this Article.
\bibitem{impact}
The impact parameter is defined here as the distance of closest approach to a given vertex
of the projection of a particle's trajectory on the plane perpendicular to the direction of the
beam.
\bibitem{rapidity}
The D0 detector utilizes a right-handed coordinate system with the
$z$ axis pointing in the direction of the proton
beam and the $y$ axis pointing upwards. The azimuthal
angle is defined in the $xy$ plane measured from the $x$
axis. The pseudorapidity is defined as $\eta \equiv  -\ln [\tan(\theta / 2)]$,
where $\theta$ is the polar angle with respect to the proton
beam direction.
\bibitem{notation}
The notation and definition of all quantities are given in Ref.~\cite{PRD}.
\end{thebibliography}
\end{document}